\documentclass[12pt,halfline,a4paper]{ouparticle}
\usepackage{breakcites}
\usepackage{textcomp}
\usepackage{xcolor}

\begin{document}

\def\lastTA#1{\textcolor{black}{#1}}

\title{Cholinergic switch between two types of slow waves in cerebral cortex}

\author{%
\name{Trang-Anh E, Nghiem\textsuperscript{[a,b,1]}, N\'uria Tort-Colet\textsuperscript{[a,2]}, Tomasz G\'orski\textsuperscript{[a,2]}, Ulisse Ferrari\textsuperscript{[c]}, Shayan Moghimyfiroozabad\textsuperscript{[a]}, Jennifer S. Goldman\textsuperscript{[a]}, Bartosz Tele\'nczuk\textsuperscript{[a]}, Cristiano Capone\textsuperscript{[a,d]}, Thierry Bal\textsuperscript{[a]}, Matteo di Volo\textsuperscript{[a,e]}, Alain Destexhe\textsuperscript{[a]}}
\address{[a] Department of Integrative and Computational Neuroscience (ICN),
Paris-Saclay Institute of Neuroscience (NeuroPSI),
Centre National de la Recherche Scientifique (CNRS), Gif-sur-Yvette, France, [b] Department of Physics, Ecole Normale Sup\'erieure, Paris, France, [c] Sorbonne Universit\'e, INSERM, CNRS, Institut de la Vision, Paris, France, [d] Istituto Nazionale di Fisica Nucleare Sezione di Roma, Rome, Italy, [e] Present address: Laboratoire de Physique Th\'eorique et Mod\'elisation, Universit\'e de Cergy-Pontoise, 95302 Cergy-Pontoise cedex, France }
\email{[1] trang-anh.nghiem@cantab.net, [2] These authors contributed equally}
}

\abstract{Sleep slow waves are known to participate in memory consolidation, yet slow waves occurring under anesthesia present no positive effects on memory. Here, we shed light onto this paradox, based on a combination of extracellular recordings \textit{in vivo}, \textit{in vitro}, and computational models.  We find two types of slow waves, based on analyzing the temporal patterns of successive slow-wave events. The first type is consistently observed in  natural slow-wave sleep, while the second is shown to be ubiquitous under anesthesia.  Network models of spiking neurons predict that the two slow wave types emerge due to a different gain on inhibitory vs excitatory cells and that different levels of spike-frequency adaptation in excitatory cells can account for dynamical distinctions between the two types. This prediction was tested \textit{in vitro} by varying adaptation strength using an agonist of acetylcholine receptors, which demonstrated a neuromodulatory switch between the two types of slow waves.  Finally, we show that the first type of slow-wave dynamics is more sensitive to external stimuli, which can explain how slow waves in sleep and anesthesia differentially affect memory consolidation, as well as provide a link between slow-wave dynamics and memory diseases.}

\date{\today}


\maketitle

\section*{Introduction}
 In both natural sleep and anesthesia, cortical dynamics are characterized by slow, irregular oscillations ($<$1 Hz) \cite{steriade1993slow}. However, certain cognitive processes, such as those involved in memory formation \cite{wilson1994reactivation,mehta2007cortico,peyrache2009replay}, are specific to deep sleep (also known as Slow-Wave Sleep, SWS), but not to anesthesia. While contrasts between sleep and anesthesia are relatively well-studied at the whole-brain scale in terms of correlations between brain regions \cite{battaglia2004hippocampal, brown2010general, AkejuBrown2017, urbain2019brain}, any underlying  differences in the microscopic dynamics of cortical neural networks and their potential mechanisms remain to be identified. In fact, while anesthesia and sleep appear to produce similar collective dynamics, it it still unknown whether any differences exist in the correlation structure of network dynamics between the two states, and how they may affect the system's ability to encode and remember stimuli. 

At the neural population level, in both sleep and anesthesia, slow oscillations emerge from the alternation between transients of high neural firing (UP states) and transients of near silence (DOWN states) \cite{steriade1993slow}. While anesthesia can present very regular UP and DOWN states \cite{niedermeyer1999burst,bruhn2000electroencephalogram,deco2009effective}, in obvious contrast with the irregular patterns observed during sleep slow waves \cite{AkejuBrown2017}, irregular regimes also exist under anesthesia for certain anesthetics and depth \cite{deco2009effective,jercog2017up,tort2019attractor}, that may appear similar to dynamics observed in natural sleep. Additionally, UP and DOWN state activity has been obtained in slice preparations \textit{in vitro} \cite{sanchez2000cellular}. Due to their general similarity in collective dynamics, slices and anesthesia, where direct pharmacological manipulation is possible, have often been used as models of natural sleep, paving the way to investigating mechanisms underlying UP and DOWN state activity. 

Nevertheless, sleep and anesthesia present different levels of neuromodulation, known to affect the neural electrophysiological properties and response profiles to external inputs. In particular, as anesthesia is characterized by lower concentrations of acetylcholine (ACh) compared to sleep \cite{jones2003arousal,McCormick1992}, resulting in enhancement of spike-frequency adaptation driven by $\mathrm{K}^{+}$ channels \cite{mccormick1989convergence}. Following the increasing electrophysiological detail available on single neuron dynamics, computational models of spiking neurons have been employed to simulate spike-frequency adaptation and account for UP and DOWN state dynamics \cite{compte2003cellular,mattia2012exploring,capone2017slow}. Indeed, comparing models with neural recordings in anesthesia has recently uncovered a mechanistic interpretation for the emergence of UP and DOWN states, where noise and neuromodulation, which controls spike-frequency adaptation, can account for the transitions between UP states and DOWN states \cite{jercog2017up}. While noise is able to trigger a transition from a DOWN to UP state, spike-frequency adaptation on excitatory cells produces a self-inhibition that, destabilizing the UP state, causes a reset to the DOWN state \cite{mattia2012exploring}. However, it is plausible that subtle mechanistic particularities at the microscopic scale could be expressed as differences in the collective network dynamics that underlie distinct computational properties of sleep and anesthesia.



In the present paper, we uncover fundamental differences in the correlation structure of slow waves during sleep and anesthesia, that can explain the different capacity for information encoding between the two states. We investigate computational models to unveil plausible mechanisms underlying these differences. These mechanisms are subsequently tested in \textit{in vitro} preparations displaying UP and DOWN state dynamics, where pharmacological manipulation is possible. In summary, we demonstrate ways in which slow-wave dynamics differ between anesthesia and slow-wave sleep, robustly across different anesthetics, brain regions, and species, and propose possible mechanisms and functional consequences of these differences.

\section*{Methods}
\section*{Neural recordings}

\subsection*{Human temporal cortex in deep sleep}
We consider the spiking activity recorded simultaneously and extra-cellularly from 92 neurons in the human temporal cortex, the same data-set used by \cite{peyrache2012spatiotemporal,dehghani2016dynamic,telenczuk2017local,le2016high,nghiem2018maximum}. The record of interest spans across approximately 12 hours, including periods of wakefulness as well as several stages of sleep. Recordings were performed in layer II/III of the middle temporal gyrus, in an epileptic patient (found to be far from the epileptic focus and not registering epileptic activity outside of generalized seizures). Data acquisition in that region was enabled by implanting a multi-electrode array, of dimensions 1 mm in thickness and 4x4 mm in area, with 96 micro-electrodes separated by 400 $\mu m$ spacings. The array was implanted to localize seizure foci. A 30-kHz sampling frequency was employed for recording. Switches in brain state (wakefulness, SWS, REM, seizure, ...) throughout the recording were noted from the patient's behavioural and physiological parameters, yielding one hour of SWS on which our analyses were focused. Spike sorting was then performed by clustering spike waveforms using automated \lastTA{expectation maximization} algorithm Klustakwik \cite{kadir2014high}, followed by manual processing with the Klusters software as described \cite{peyrache2012spatiotemporal}. Using spike sorting methods on the obtained data, 92 neurons were identified; we only retained neurons spiking throughout the recording for our analyses, amounting to 71 neurons of which 21 were I neurons. Analysis of the spike waveforms for each of these neurons allowed their classification as putative excitatory (E) and inhibitory (I) neurons. Using the spike times of each neuron, cross-correlograms for all pairs of neurons were also computed to determine whether each neuron's spikes had an excitatory (positive correlation) or an inhibitory (negative correlation) effect on other neurons through putative monosynaptic connections. It should be noted that neurons found to be excitatory exactly corresponded to those classified as regular spiking (RS), while all inhibitory neurons were also fast spiking (FS). Spikes were binned into 1 ms wide time bins for all subsequent analyses. A statistical analysis of UP DOWN states duration is presented in Fig.S\ref{figS1_data} (panels A-D and E left).

\subsection*{Monkey premotor cortex in deep sleep}
Spiking activity (the same data-set as used in \cite{dehghani2012avalanche,dehghani2016dynamic,telenczuk2017local,le2016high,nghiem2018maximum}) in layer III/IV of the premotor cortex of a macaque monkey was recorded by multi-electrode arrays throughout a night. A 10-kHz sampling frequency was employed for recording. Spiking events were detected when signals crossed a threshold above the noise band, and spike-sorted using Offline Sorter (Plexon, Inc, Dallas, TX, USA) as described in \cite{dehghani2012avalanche}. Classification of brain states, for extraction of SWS periods, was performed by visual inspection of the Local Field Potential (LFP), over time periods of 5 s, by identifying as SWS periods presenting large-amplitude oscillations in the 1-2 Hz frequency range \cite{nghiem2018maximum}, of which 141 spiked throughout the whole recording, yielding three hours of SWS  data  for subsequent analyses. All analyses in this work were performed with spikes binned into 1 ms time bins.

\subsection*{Rat prefrontal cortex in deep sleep}
The analysis was performed on the dataset of single unit activities  previously employed in \cite{peyrache2009replay,benchenane2010coherent,tavoni2017functional}.
Here we provide a short description only.
Five Long-Evans male rats were chronically implanted with tetrodes in the prelimbic subdivision of the medial prefrontal cortex and in the intermediate-ventral hippocampus. Tetrodes in the hippocampus were used for identification of non-REM sleep periods, through a clustering analysis of the LFP power within the cortical delta band ($1-4$ Hz), hippocampal theta ($5-10$ Hz) and cortical spindles ($10-20$ Hz) together with estimates of the speed of head movements. Tetrodes in the cortex were used for single unit recording. Spike sorting has been performed using \textit{KlustaKwik} \cite{kadir2014high}.

Recordings were organized in daily sessions, where the rat undergoes a first sleeping epoch, then a task learning epoch, in which the rat performs an attentional set shift task on a Y shaped maze, and finally a second sleeping epoch, each epoch lasting 30 mins. In general, the neurons recorded differed \lastTA{from one session} to another, with the number of cells recorded per session varying between $10$ and $50$. For the analysis of UP and DOWN state duration, the results from all \lastTA{sessions} from the same rat were joined together, but pre-task and post-task sleep were kept separated.

\subsection*{Monkey primary visual cortex under sufentanil anesthesia}
The data-set may be found at \cite{kohn2016utah}, as described in \cite{smith2008spatial}. Four adult macaque monkeys were recorded using a total of six multielectrode arrays implanted in the primary visual cortex. Sufentanil (4-18 $\mu$g/kg/hr) was used for anesthesia. Recordings were obtained while animals viewed a uniform gray screen, over periods of between 20 and 30 minutes long. Spontaneous spiking activity from $~$70 – 100 neurons was recorded and spike-sorted for each array. Spike sorting was performed \cite{smith2008spatial} using an automated algorithm for clustering spike waveforms based on mixtures of multivariate distributions \cite{shoham2003robust}. Spikes were binned into 1 ms time bins for subsequent analyses. A statistical analysis of UP DOWN states duration is presented in Fig.\ref{figS1_data}E right. 

\subsection*{Rat primary visual cortex under ketamine anesthesia}
7 adult male Wistar rats \lastTA{weighing} 211 $\pm$ 58 g (mean $\pm$ s.d.) were anesthetized via intraperitoneal injection of ketamine (120 mg/kg) and medetomidine (0.5 mg/kg). Atropine (0.05 mg/kg) was injected subcutaneously to prevent respiratory secretions. Rectal temperature was maintained at 37\textdegree C. A craniotomy was performed to access the primary visual (V1) cortex (7.3 mm AP, 3.5 mm ML) of the right hemisphere \cite{PaxinosWatson2004}. 

Recordings of cortical activity under anesthesia were obtained with a 16-channel silicon probe (1 shank with 16 linearly spaced sites at $100 \mu m$ increments with impedances of $0.6-1 M\Omega$ at 1kHz (NeuroNexus Technologies, Ann Arbor, MI)) introduced perpendicularly in V1 under visual guidance until the most superficial recording site was aligned with the cortical surface. Signals were amplified (Multi Channel Systems), digitized at 10kHz and acquired with a CED acquisition board and Spike 2 software (Cambridge Electronic Design, UK). Recordings had an average length of 951.46 +/-219.30 seconds. 
UP and DOWN states were singled out by thresholding the multi-unit activity (MUA), which was estimated as the power of the Fourier components at high frequencies  (200-1500 Hz) of the extracellular recordings (LFP) \cite{Reig2010, sanchez2010inhibitory, RuizMejias2011,mattia2016metastable,tort2019attractor}. For each experiment, we selected the channel with maximum MUA during the UP state, which corresponds to cortical layer 5 \cite{tort2019attractor,mattia2016metastable}.

All experiments were supervised and approved by the local Ethics Committee and were carried out in accordance with the present laws of animal care, EU guidelines on protection of vertebrates used for experimentation (Strasbourg 3/18/1986) and the local law of animal care established by the Generalitat of Catalonia (Decree 214/97, 20 July). 

\subsection*{Cat parietal cortex in sleep and ketamine anesthesia}
Two adult cats weighting
$2.5-3.5$ kg \lastTA{were} chronically implanted eight electrodes each, in the parietal cortex (Brodmann areas 5-7). One cat was recorded during natural sleep \cite{destexhe2009self}, while the other was recorded under ketamine-xylazine anesthesia (2-3 mg/kg, intramuscular), with additional doses of anaesthetic (3-7 mg/kg) administered as necessary (2-3 times during the experiment) \cite{contreras1995cellular}. In bipolar recordings, the polarity was adjusted such that the sharp potentials of slow-wave complexes are negative. Signals were recorded on an eight-channel digital recorder (Instrutech, Mineola, NY) with an internal sampling rate of 11.8 kHz per channel and four-pole Bessel filters. The data was digitized off-line at 250 Hz using the Igor software package to obtain LFP. 

\subsection*{Mouse entorhinal cortex slice preparations}
We prepared brain slices exhibiting spontaneous slow waves in entorhinal cortex using a method described in \cite{2012-tahvildari/mccormick}. The mice were of wild-type (C57BL/6J) and 11-18 days old.

The dissection and slice cutting were performed in an ice-cold cutting solution containing (in mM): 85 NaCl, 75 sucrose, 3 KCl, 26 NaHCO\textsubscript{3}, 1.25 NaH\textsubscript{2}PO\textsubscript{4},
3.5 MgSO\textsubscript{4}, 0.5 CaCl\textsubscript{2}, 10 glucose, 3 myo-inositol, 3 Na-pyruvate, 0.5 L-ascorbic acid and aerated with 95\% O\textsubscript{2} and 5\% CO\textsubscript{2}.
Lower concentrations of Na\textsuperscript{+} and Ca\textsuperscript{2+}, and a higher concentration of Mg\textsuperscript{2+} in the cutting solution, compared to a standard ACSF, are applied to minimize neuronal damage during cutting.

We cut slices at a 15\textdegree angle off the horizontal plane with the thickness of 310 $\mu$m. After cutting, slices were placed in a cutting solution at temperature of 35\textdegree  C for 30 min.

The slices were then kept at room temperature in a storing solution containing (in mM):
126 NaCl, 3 KCl, 26 NaHCO\textsubscript{3}, 1.25 NaH\textsubscript{2}PO\textsubscript{4},
2 MgSO\textsubscript{4}, 2 CaCl\textsubscript{2}, 10 glucose, 3 myo-inositol, 3 Na-pyruvate, 0.5 L-ascorbic acid.

For recording, the slices were transfered to a submersion chamber and placed between nylon nets. The well-oxygenated recording solution was flowing with the speed of 4ml/min.
The recording solution was similar to the storing solution, with only CaCl\textsubscript{2} and MgSO\textsubscript{4} concentrations reduced to 1.2 and 1 mM respectively. The extracellular electric field was recorded with glass electrodes with a resistance of 2-3 $M\Omega$.
The electrode was placed in layer 2/3 of the entorhinal cortex.

Electrophysiological data was acquired using the ELPHY software \cite{sadocELPHY}. The multi-unit activity was obtained from the signal by calculating the time-averaged power of the signal in the frequency range (0.3 - 2 kHz).  

\section*{Spiking network model}

We consider a population of $N=10^4$ neurons connected over a random directed network with probability of connection $p=5\%$. The population comprises excitatory and inhibitory neurons, with 20\% inhibitory neurons.
The dynamics of each of the neuron types is based on the adaptive integrate and fire model, described by the following equations

\begin{equation}\label{microscopic_eq}
c_{m} \frac{dv_i}{dt} = g_L(E_L-v_i)+g_L k_a e^{\frac{v_i-v_{thr}}{k_a }} -w_i + I_{syn} + \sigma\xi_i(t)
\end{equation}

\begin{equation}\label{microscopic_eq_adap}
\frac{dw_i}{dt}= -\frac{w_i}{\tau_w} +b\sum_{t_{sp}(i)}\delta(t-t_{sp}(i)) +a(v_i-E_L),
\end{equation}

where $c_{m}$ is the membrane capacity,  $v_i$ is the voltage of neuron $i$ and whenever $v_i>v_{thr}$ at time $t_{sp}(i)$ , $v_i$ is reset to the resting voltage $v_{rest}$ and fixed to that value for a refractory time $\tau_{r}$.
The exponential term mimics activation of sodium channels and parameter $k_a$ describes its sharpness. Inhibitory neurons are modeled according to physiological insights \cite{destexhe2009self} as fast spiking (FS) neurons with no adaptation while the strength $b$ of spike-frequency adaptation in excitatory regular spiking (RS) neurons is varied.
The synaptic current $I_{syn}$ received by neuron $i$ results from the spiking activity of all pre-synaptic neurons $j\in\mathrm{pre}(i)$ of neuron $i$. This current can be decomposed \lastTA{as the result of input received} from excitatory E and inhibitory I pre-synaptic spikes  $I_{syn}=(E_e-v_i)I^e_{syn}+ (E_I-v_i)I^I_{syn}$. Notice that we consider voltage dependent conductances.
Finally, we model $I^x_{syn}$ as a decaying exponential function that takes kicks of amount $Q_x$ at each pre-synaptic spike, i.e.:

\begin{equation}
I^x_{syn}(t) = Q_x\sum_{exc. pre}\delta(t-t^x_{sp}(i))e^ {-\frac{t-t^x_{sp}(i)}{\tau_x }},
\end{equation}

where $x$ represents the population type ($x=E,I$), $\tau_x $ is the synaptic decay time scale  and $Q_x$ the quantal conductance. We have the same equation with $E\rightarrow I$ for inhibitory neurons. Every neuron $i$ receives a uniformly distributed white noise $\xi(t)$ of zero mean and instantaneously decaying autocorrelation $\langle\xi_i \rangle=0$, $\langle\xi_i(t)\xi_j(t+s) \rangle=\delta_{i,j}\delta(t-s)$. The noise amplitude $\sigma$ is a piecewise constant function of time, i.e. its value stays constant for a time window of length $T$ and is extracted from a uniform distribution of amplitude $\Delta$. In our simulations $\Delta$ varies and we use $T=100$s, in accordance to the observed variability of UP and DOWN states duration during sleep.

A transient of 1s after simulation onset is discarded from all analyses. An example of the dynamics of this system is reported in Fig.\ref{fig1_model}A and statistics of UP and DOWN states durations obtained with the model are shown in Fig.\ref{fig1_model}B-E.

To deliver a stimulus to the network, each neuron receives an external Poissonian spike train of frequency 0.05 Hz for a duration of 50 ms. Stimuli are delivered halfway through the first UP state after the discarded transient. To directly compare network dynamics in the presence and absence of a stimulus, the network connectivity matrix and initial conditions are the same in both simulations, such that dynamics before the stimulus onset are identical, and differences in dynamics following the onset are only due to the stimulation. The cumulative spike count is computed in each case, at each point in time. The normalized distance between the spike counts with and without stimulus is defined by:
\begin{equation}
D = \frac{|s' - s|}{<s>},
\end{equation}
where $s$ is the spike count for the unstimulated network, $s'$ is the spike count for the stimulated network, and $<\cdot>$ denotes time averaging. Each simulation lasts 30s, and the procedure is repeated 50 times, each time with different connectivity realization and initial condition set. In Fig. \ref{fig4_sensitivity} we report the average value of $D$ over different realizations, together with its standard deviation.

\section*{UP and DOWN state detection and statistics}
In all recordings, only time intervals displaying slow wave dynamics were used in subsequent analyses. Other stages in sleep recordings, such as light sleep and Rapid Eye Movement (REM) sleep, were excluded from the analysis.

The method to detect UP states from unit activity (\cite{renart2010asynchronous}, Section $1.3.3.$ of Suppl Mat.) considers the sum of all cells' spike trains (bin size of 1 ms), $K(t)=\sum_i \sigma_i(t)$.
The instantaneous population activity $m(t)$ is the smoothed $K(t)$, by convolution with a Gaussian density with width $\alpha = 10$ ms. Any period of time for which the instantaneous population activity\\ $m(t) > \theta\max(m(t))$ is considered an UP state, where the threshold $\theta$ was chosen in terms of the sparseness and non-stationarity of each data-set. $\theta = 0.2$ was for most data-sets, as in \cite{renart2010asynchronous}, except human SWS and the spiking model, where $\theta = 0.02$, and slice preparations, for which $\theta = 0.5$. Note that the threshold was chosen to be low due to UP states themselves being relatively quiescent in this data-set (human temporal cortex, layer 2-3, recorded with multi-electrode array). Indeed, as estimated from the spikes of the recorded $10^2$ neurons the firing rate is low even in the UP state. While UP states are usually visible by eye (Fig. 1A), a low threshold is needed to discriminate UP states from DOWN states. States lasting less than $50$ ms were excluded by considering it a part of the previous sufficiently long state. States longer than $5$ s are discarded from the analysis. Parameters used for detection were determined by visual inspection of the detection quality. It was also verified that slight variation of these parameters did not qualitatively affect the results presented in this work (Fig. \ref{figS3_detection}), suggesting observed effects are robust rather than artefacts of the detection method. 

This method was tested against a different method where UP and DOWN states were singled out by setting a minimum state duration of 80ms and a threshold in log(MUA) values at 1/3 of the interval between the peaks of the bimodal distribution of log(MUA) corresponding to the UP and DOWN states. The algorithm,  adapted from \cite{mattia2012exploring,RuizMejias2011, tort2018bimodality, tort2019attractor}, yielded qualitatively identical results.  

To detect UP and DOWN states from LFP, a similar approach was applied, by setting a threshold three standard deviations away from the mean LFP signal. The points of signal crossing from above to below the threshold were labeled DOWN to UP state transitions. Conversely, the closest local minimum preceding a crossing from below to above was identified as an UP to DOWN transition. Therefore, periods of high LFP were marked as DOWN states, and periods of low LFP as UP states. States shorter than $60$ ms were considered part of the previous state lasting $60$ ms or longer.

The Pearson correlation was then employed to evaluate the strength and significance of the correlation between UP state and DOWN state durations. As a further test for significance, the information present in time structure was destroyed by shuffling all DOWN state durations, while leaving UP state durations in their empirical order, and computing the Pearson correlation again. This procedure is repeated 1,000 times, and the mean and standard deviation of the Pearson correlations obtained each time are calculated. The interval contained within two standard deviations above and below the mean of correlations obtained from shuffled is considered as a confidence interval. Indeed, a correlation well outside of this interval is highly unlikely to have been produced by a chance arrangement of UP and DOWN states in time, given the empirical distribution of their durations, and implies a non-trivial structure in time. 

This procedure is used to evaluate the correlation between each UP state and the DOWN states surrounding it, $C(D_{n+k}, U_n)$, with $k = 0$ denoting the previous DOWN state to the considered UP state, negative $k$ denoting previous DOWN states more distant in time, and positive $k$ denoting DOWN states following the UP state of interest.

The scripts used for UP and DOWN state detection and analysis can be found at \url{https://github.com/tnghiem/UP and DOWN-analysis}.

\section*{Results}
\subsection*{Two types of slow waves found in anesthesia and natural sleep}

\begin{figure}[!htb]
\centering
\includegraphics[width=.65\linewidth]{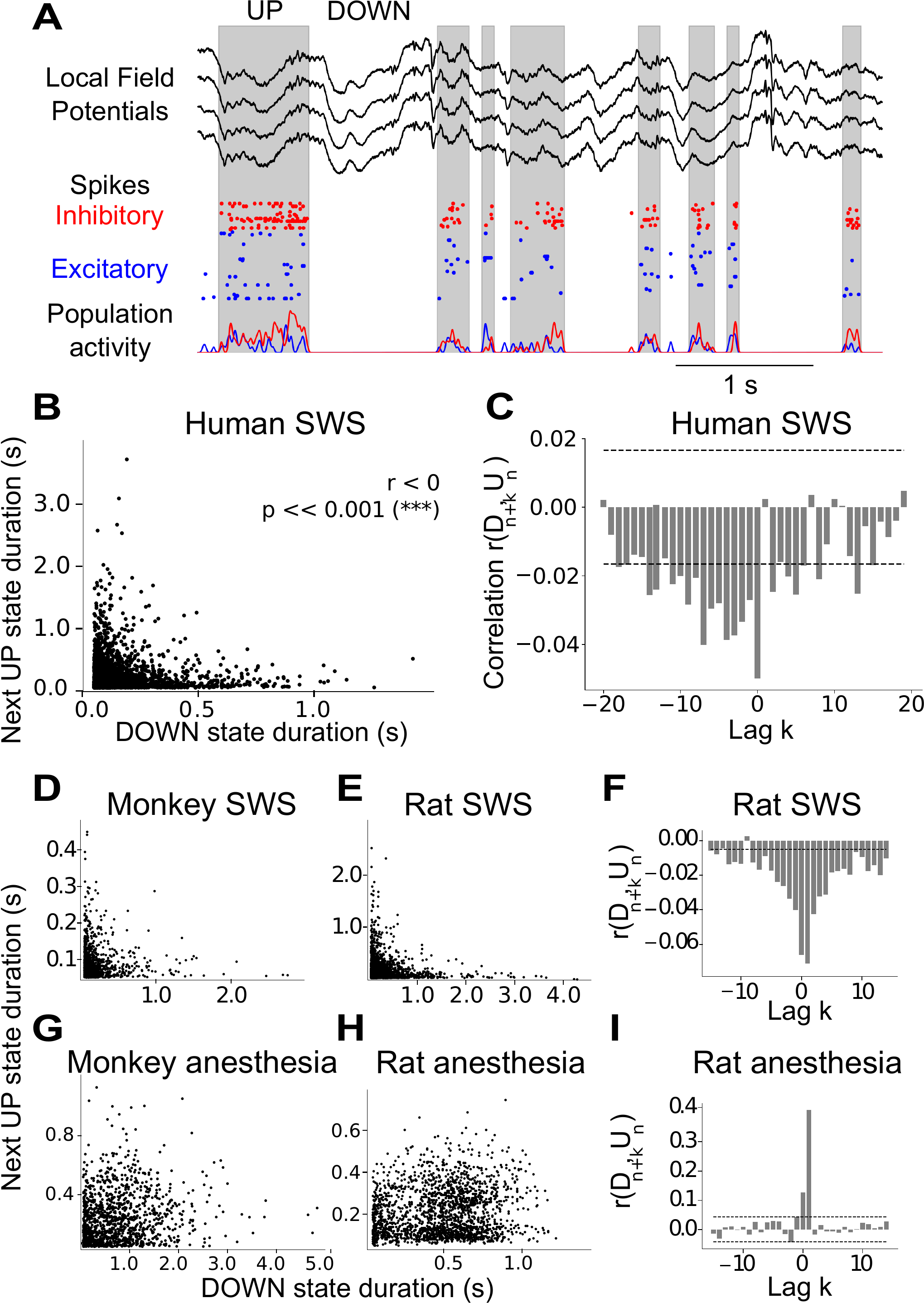}
\caption{\textbf{Different types of slow waves during sleep and anesthesia, across species, brain regions, and anesthetics.} (A) LFP (top) and spiking data (bottom) recorded by multi-electrode array implanted into a human patient's temporal cortex. Slow oscillations ($<$1 Hz) visible in the LFP correspond to an alternation between transients of high and low firing rate, i.e. UP and DOWN state dynamics, evident in the spiking activity (gray: UP state detection based on population spike count, see Methods). (B) UP state duration against previous DOWN state duration, showing a clear negative correlation. (C) Bar plot of Pearson correlation $r(D_{n+ k}, U_n)$ as a function of lag $k$. Two standard deviations of the Pearson correlations when shuffling state durations (dashed lines) provide an interval of confidence outside of which empirical correlations may be considered non-trivial. The same analyses are reported for other species during sleep and anesthesia:  sleep in the monkey premotor cortex (D) and in the rat pre-frontal cortex (E),  anesthesia in the monkey (G) and rat V1 (H). Panels (F) and (I) report the lag-correlation during sleep and anesthesia in the rat.}
\label{fig1_data}
\end{figure}

To compare with previous anesthesia results \cite{jercog2017up,deco2009effective}, 
we consider the activity of a population of $~10^2$ neurons recorded from the temporal cortex of a human patient (Fig. \ref{fig1_data}A-C) during sleep. The dynamics are characterized by slow waves, as evident from LFP associated with an alternation of low (DOWN) and high (UP) activity periods, as visible in spiking activity. In all our analyses UP and DOWN states are defined based on neuron spiking activity (see Methods). The time duration of both DOWN and UP states are variable, following an exponential, long-tailed distribution (Fig. \ref{figS1_data}), similar to what has been reported for anesthesia recordings \cite{jercog2017up}.

While the distributions of UP and DOWN state durations found in sleep and anesthesia are similar, surprisingly, the temporal distribution of UP and DOWN state durations is different. In anesthesia, DOWN state and next UP state durations are positively correlated, in other words long DOWN states are followed by long UP states, as shown in previous work \cite{jercog2017up}. However, we find that in sleep, DOWN state and next UP state durations are negatively correlated, as long DOWN states are followed by short UP states (Fig. \ref{fig1_data}B). Indeed, during sleep, while long UP states can occur after short DOWN states, UP states following long DOWN states are consistently short.

Because a difference in temporal correlation was found for adjacent UP and DOWN epochs between sleep and anesthesia, we next explored whether the network retains a memory of previous epochs further in time. To this end, correlations between the n-th DOWN state and the (n+k)-th UP state duration were explored. Here, $k = 0$ denotes the UP state following the $n$th DOWN state, as studied so far, while $k = -1$ denotes the UP state preceding the $n$th DOWN state in time. As shown in Fig.\ref{fig1_data}C, the length of time correlations remain between DOWN and UP states lag $k$  remains significantly negative up to a separation on the order of five slow-wave cycles (each cycle consisting of one UP and consecutive DOWN state) in sleep data. In contrast,  the correlations at lag $k$ decay to zero immediately after one slow-wave cycle in anesthesia \cite{jercog2017up}. Thus, the network retains a memory of previous cycles significantly longer in SWS than anesthesia. 

In order to investigate whether the correlation between UP and DOWN state duration and its memory through time are specific to brain states across species and regions of cortex, data from V1 of animals under different anesthetics (monkey under sufentanil, rat under ketamine and medetomidine), and several animals sleeping (human temporal cortex, monkey premotor cortex, and rat prefrontal cortex) were analyzed. 

The results of sleep analyses are reported in Fig.\ref{fig1_data} D-I scatter plots of DOWN versus consecutive UP state duration, showing negatively correlated banana-shaped distributions (for sleeping rats, $5/5$ of animals).


Conversely, in anesthetized recordings (Fig. \ref{fig1_data}G,H), results are consistent with previously published results (Fig. 2C from \cite{jercog2017up}), showing positive or non-significant correlations (for anesthetized monkeys, 5/6 recordings showed significant correlation, 1/6 non-significant correlation; for anesthetized rats, 4/7 animals showed a significant positive correlation, 2/7 showed a positive non-significant correlation and the remaining 1/7 showed a negative significant correlation).

Moreover, the structure of correlations between the duration of non-consecutive states, across multiple lags, is also robustly distinct between sleep and anesthesia. When comparing within the same species (rat, Fig. \ref{fig1_data}F,I), correlations remained significant across many more slow-wave cycles during sleep (consistent with human sleep, Fig. 1C) than anesthesia. 


In spiking data, correlation between UP and DOWN state duration are negative during sleep, but positive or non-significant during anesthesia across a variety of animals and cortical regions. Consistent with these observations, LFP signals recorded in cat parietal cortex (Brodmann areas 5-7) show negative correlation between consecutive DOWN and UP state duration during natural sleep, but positive correlation during anesthesia \ref{figS4_LFPCat}. These results highlight that both spiking activity (Fig. \ref{fig1_data}) and LFP (Fig. \ref{figS4_LFPCat}) reveal clear differences \lastTA{in the correlation} structure of network dynamics during sleep versus anesthesia.


\subsection*{Existing spiking network model reproduces anesthesia- but not sleep-like slow waves}
In order to investigate the mechanisms underlying the UP and DOWN state duration correlations, we use a network of spiking neurons with conductance-based (COBA) synapses. The network is composed of 80\% RS (regular spiking) excitatory and 20\% fast spiking (FS) inhibitory neurons. Every neuron is modeled as an Adaptive Exponential integrate-and-fire cell (AdExp) \cite{brette2005adaptive}. In the absence of adaptation, the system is characterized by two stable states: a near-silent state (DOWN state) and a relatively high-activity state (UP state).

To allow for transitions between the two states, every neuron receives an independent identically distributed (i.i.d.) zero-mean noise of amplitude $ \sigma$ that permits a jump from the DOWN to the UP state. The presence of spike-frequency adaptation of strength $b$ (see Methods) for RS neurons \cite{pospischil2008minimal} allows the system to transition back to the DOWN state. Indeed, in RS neurons adaptation builds up as they spike, i.e. during UP states, and consequently reduces the firing rate of the excitatory population, which can cause the transition to a DOWN state. Adaptation decays exponentially throughout time when the neuron is silent, for instance during DOWN states (see Methods for equations). 

\begin{figure}[!htb]
\centering
\includegraphics[width=.85\linewidth]{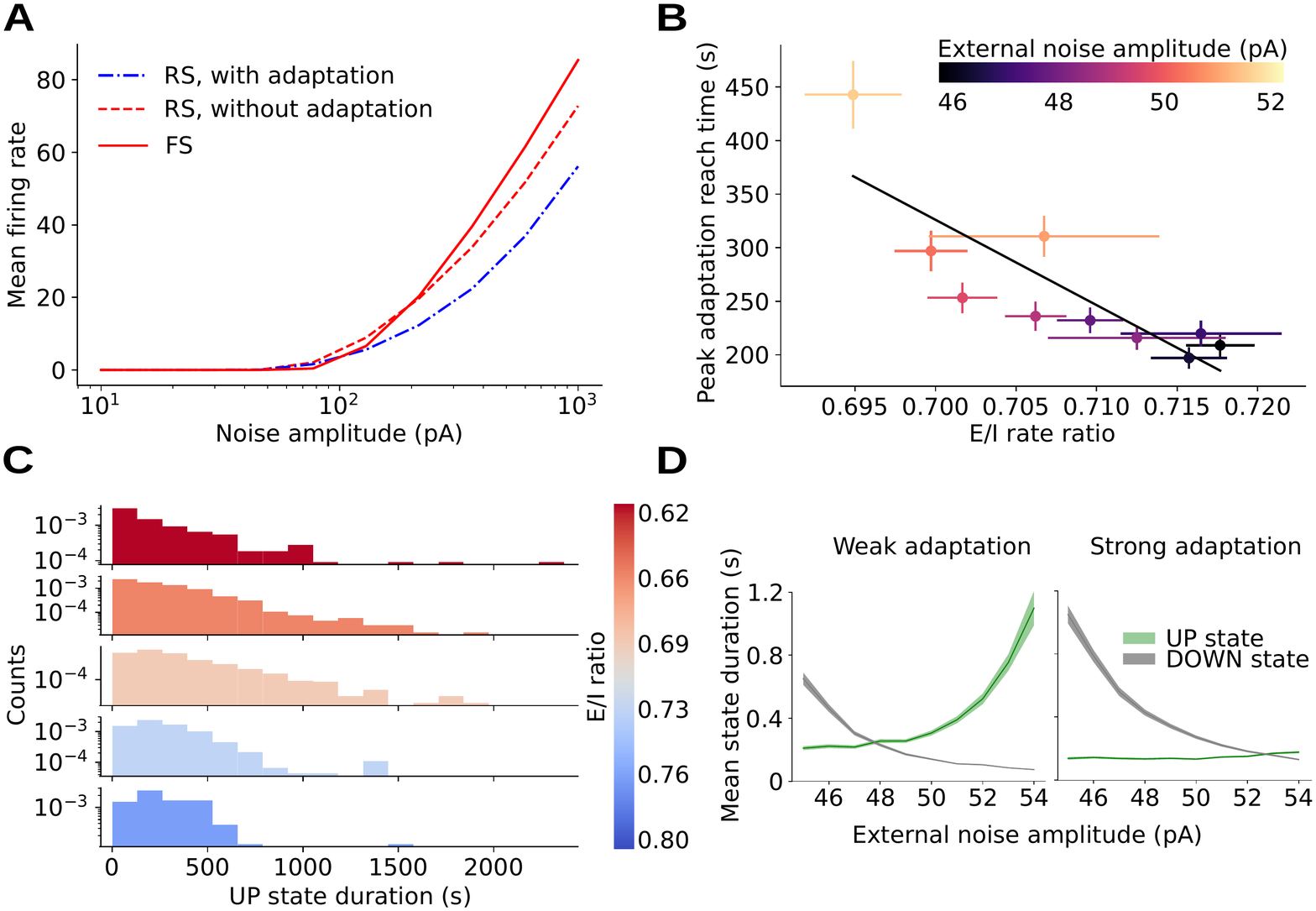}
\caption{\textbf{Enhanced noise sensitivity in inhibitory neurons as a mechanism for prolonged UP states.}  (A) Firing rate of isolated E (regular-spiking, blue), I cells (fast-spiking, red) cells as a function of input noise amplitude, and ersatz I cells with the same single-cell parameters as E cells (see Supplementary Table 1 for values) and zero adaptation, such that the gain is similar to that of E cells (red, dotted). Inhibitory cells are observed to have a higher gain at high noise amplitude values. (B) Time taken to reach peak adaptation value in UP state against mean ratio of E to I firing rate at the onset of UP states for varying amplitude of external noise (error bar: standard error of the mean). (C) Histogram of UP state duration for E/I ration in different intervals (different colors). UP states can reach longer duration for low E/I ratio. Statistics gathered for UP states over simulations with different values of noise amplitude ($ 46 \mathrm{pA} < \sigma < 52 \mathrm{pA}$). (D) Mean UP and DOWN state duration vs amplitude of external noise (shaded: standard error of the mean), for low (b = 10 nS, left) and high (b = 50 nS, right) adaptation strength in the spiking network model (see text and Methods).}
\label{fig2_EI}
\end{figure}

\begin{figure}[!htb]
\centering
\includegraphics[width=.65\linewidth]{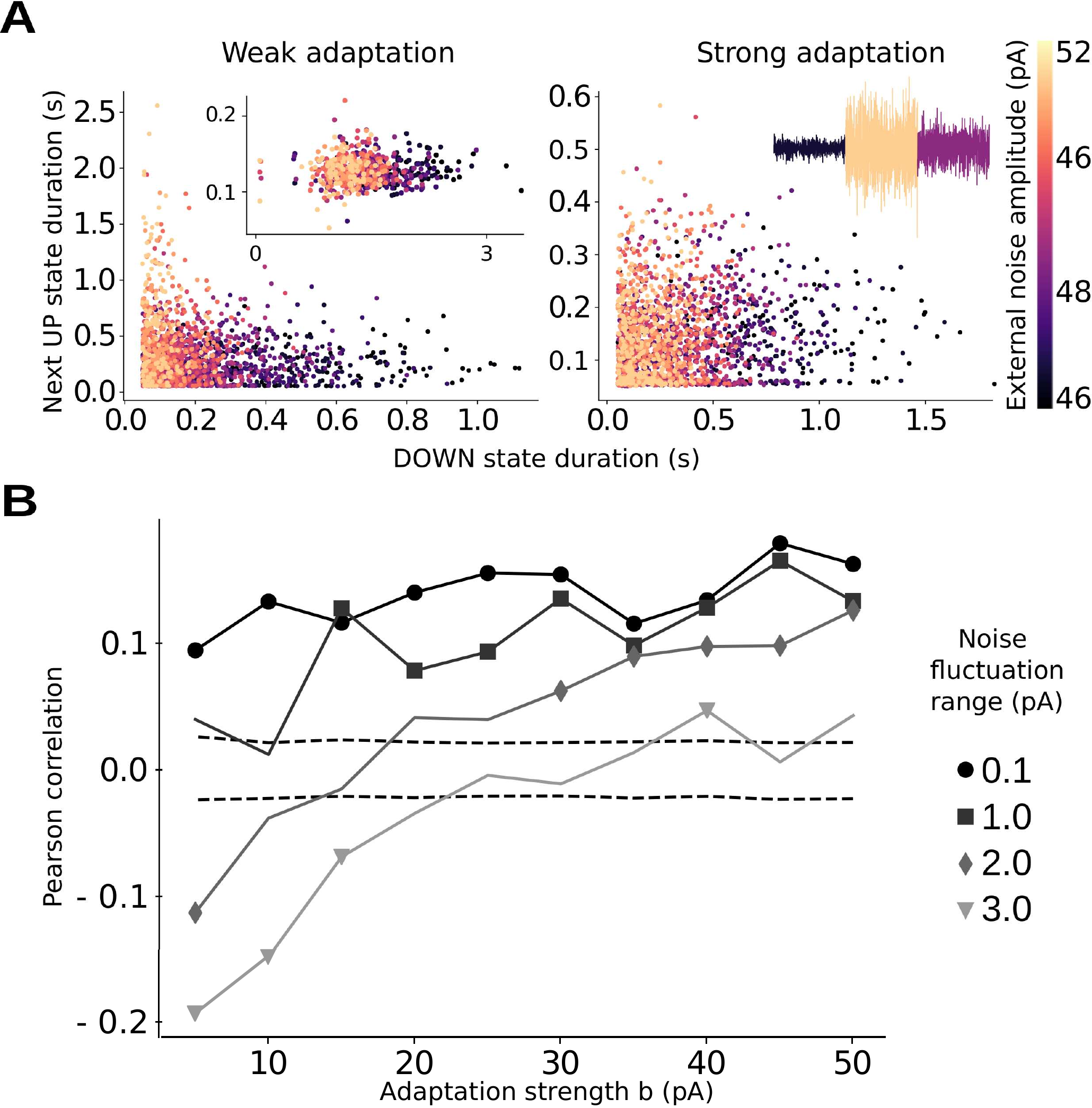}
\caption{\textbf{Interplay between spike-frequency adaptation strength and noise amplitude allows for a transition from sleep-like to anesthesia-like slow waves in a spiking network model.}  (A) UP state duration against previous DOWN state duration for external noise amplitude varying throughout time in the range $[-\Delta,\Delta]$. Here, $\Delta = 3$ pA, showing negatively correlated (r = -0.15, p $<<$ 0.001), `sleep-like' durations for low adaptation (b = 10 nS, left) and uncorrelated (r = -0.01, p $>$ 0.05), `anesthesia-like' durations for high adaptation (b = 50 nS, right). Inset left: replacing I cells with `ersatz inhibitory' cells endowed with the same physiological properties as E cells (sodium sharpness and leakage reversal, see Supplementary Table 1 for values) such that the gain is similar to that of E cells (see Fig. 2A, dashed line). UP state durations are reduced and the negative correlation vanishes (r = -0.02, p $>$ 0.05). Inset right: sketch of noise time course, in which different colors represent different noise amplitude $\sigma$. (B) Variation of next UP to previous DOWN state correlation as a function of adaptation strength, for different ranges $\Delta$  of noise fluctuation (markers: significant correlations, dashed lines: confidence interval obtained by shuffling, see Methods).}
\label{fig3_model}
\end{figure}

Such mechanisms for the emergence of UP and DOWN state dynamics have been so far established in the literature (see e.g. \cite{jercog2017up,holcman2006emergence}). We observed the same mechanism in a spiking network model of Adaptive integrate and fire neurons  with voltage-dependent synapses and a different gain between excitatory and inhibitory cells. The network  dynamics is characterised by an alternation between UP and DOWN states whose durations follow an exponential distribution, in accordance with experimental data (see Fig \ref{fig1_data}C for data and Supplementary material for the model). 

For a fixed value of noise amplitude $\sigma$ we observe a positive correlation between UP and DOWN state duration, where adaptation strength $b$ changes UP and DOWN state length with no obvious effect on their correlation. The correlation between UP and DOWN state duration remains  positive or non-significant over the all range of $b$ values here investigated. This is consistent with adaptation having decayed after long DOWN states: following noise-triggered onset,  the following UP state displays a high rate of activity that may sustain a long UP state. Consequently, long DOWN states tend to be followed by long UP states, hence the positive correlation. Exploring the parameter space by varying other parameters such as neural excitability or synaptic quantal conductances,  positive or non-significant correlations were also always obtained.

In accordance with previously reported results \cite{jercog2017up}, the model discussed so far is suitable for UP and DOWN state dynamics during anesthesia but not for sleep, where we have shown a clear and robust negative correlation. This shows additional elements are needed to accurately model the empirical UP and DOWN state dynamics during sleep.  

\subsection*{High inhibitory neuron gain and adaptation gating of fluctuations needed for switching between anesthesia-like and sleep-like dynamics}
In order to investigate mechanisms entering into play during sleep we observe that, additionally to adaptation strength, another natural  parameter for affecting UP and DOWN state duration is the amount of noise $\sigma$. 
The role of external noise in our system is complex as noise affects excitatory and inhibitory cells differently, thus rendering the collective scenario non-trivial. In fact, in Fig. \ref{fig2_EI}A isolated inhibitory FS cells are shown to have a higher gain in their response to external noise amplitude $\sigma$ with respect to excitatory RS cells. As a result, we expect that the higher the noise amplitude $\sigma$, the lower the ratio of excitatory to inhibitory activity (E/I ratio). In fact, by performing network simulations with varying $\sigma$ we observe that the E/I ratio at UP state onset (where adaptation has not yet built up such that dynamics are noise-driven) decreases with $\sigma$ (Fig. \ref{fig2_EI}B). This effect has crucial consequences on population dynamics in terms of UP state duration. In fact, as shown in Fig. \ref{fig2_EI}C, UP state duration is enhanced and can reach especially high values for low E/I ratio. In particular, for low E/I ratio, large inhibition slows the increase of excitatory activity at the onset of UP states. Therefore decreasing the E/I ratio at UP state onset results in decelerating subsequent adaptation build throughout the UP state, which can be quantified the time taken before the maximum value of adaptation is reached in each UP state (Fig. \ref{fig2_EI}B), thus permitting longer UP states. As a result, high noise amplitude $\sigma$ implies long UP states as reported in simulation results (Fig.\ref{fig2_EI}D). 
In our model, UP to DOWN state transitions are driven by deterministic adaptation build-up rather than by stochastic noise fluctuations. Therefore, due to the higher gain of inhibitory FS cells, the higher the noise amplitude the longer UP states can be. Additionally, since increased noise amplitude causes larger fluctuations more likely to occasion DOWN to UP state transitions, the higher the noise amplitude the shorter DOWN states tend to be. This implies that UP and DOWN state durations vary in an anti-correlated fashion with noise amplitude $\sigma$. In other terms, if $\sigma$ were to vary throughout time, a negative correlation could be observed between consecutive DOWN and UP state durations.

To account for correlation inversion in our model, we introduce a parameter, $\Delta$, describing the variability of noise amplitude $\sigma$ throughout time. Here, the noise $\sigma$ takes successive values within a range $\Delta$ of amplitudes $\sigma$ (see Methods), where each value is held constant over a time interval of duration $T$ (inset of Fig.\ref{fig3_model}A, right). It should be noted that the resulting UP and DOWN duration correlation does not depend on the specific choice of $T$, as far as it is long enough to contain a sufficient number of UP and DOWN state transitions in order to obtain well-defined UP and DOWN state statistics (in the plots shown in Fig \ref{fig3_model} $T = 100$s).

By introducing variation of noise amplitude in time, a banana shape is observed in the scatter plot of UP and DOWN state duration (see Fig.\ref{fig3_model}A, left). Accordingly, a negative correlation between UP and DOWN state durations emerges increasing the range $\Delta$ of variation of the noise amplitude $\sigma$. In order to further investigate the mechanism based on a higher gain of inhibitory FS cells, we performed a simulation in a set-up where excitatory and inhibitory neurons have similar gains (see red dashed line in new Fig.\ref{fig2_EI}A). We demonstrate that negative correlation between DOWN and next UP state durations require a higher gain on FS-I cells in the model. This is simulated by endowing both E and I cells with RS properties (sodium sharpness and leakage reversal, see Supplementary Table 1 for parameter values) hence lowering the gain of I cells to match that of E cells, while keeping all other parameters identical to those used to produce Fig 2AB, left. As shown below on the left panel, this yields consistently shorter UP states for all noise amplitudes (green), verifying the role played by FS-I cells in sustaining UP states. Moreover, in this condition DOWN and next UP state durations are found to be uncorrelated (right panel, Pearson correlation r = 0.03, p = 0.5), which confirms that higher FS gain is a necessary mechanism to obtain sleep-like negative correlation by varying noise amplitude (Fig. \ref{fig3_model}A inset, right). 

Additionally, a fundamental role for the emergence of sleep-like negative correlation is played by spike frequency adaptation. Indeed, the inverse correlation disappears as the adaptation strength $b$ is increased (Fig. \ref{fig3_model}B). Accordingly, a negative correlation between UP and DOWN state durations emerges increasing the range $\Delta$ of variation of the noise amplitude $\sigma$. Moreover, for sufficiently high $\Delta$, an increase in adaptation strength $b$ is able to induce a transition from negative (sleep-like) to positive (anesthesia-like) correlation. In other words, adaptation is able to filter out noise variability, thus determining a positive correlation. 

Apparent in the scatter plots of panel \ref{fig3_model}B, when adaptation is low, various values of noise amplitude $\sigma$ (indicated by colors) cluster together in the scatter plot, altogether yielding a banana shape. Conversely, for high adaptation strength, data representing different values of noise amplitude overlap in the scatter plot, resulting in a non-significant or positive correlation. This can be understood as strong adaptation limiting the duration of  UP states (green line in Fig. \ref{fig3_model}A, right), even in the presence of strong noise, and more generally controlling the transitions between UP and DOWN states. In sum, the model highlights the dominant mechanisms at work in each brain state, with the system being strongly adaptation driven in anesthesia, and more fluctuation driven in sleep.

\begin{figure}
\centering
\includegraphics[width=.85\linewidth]{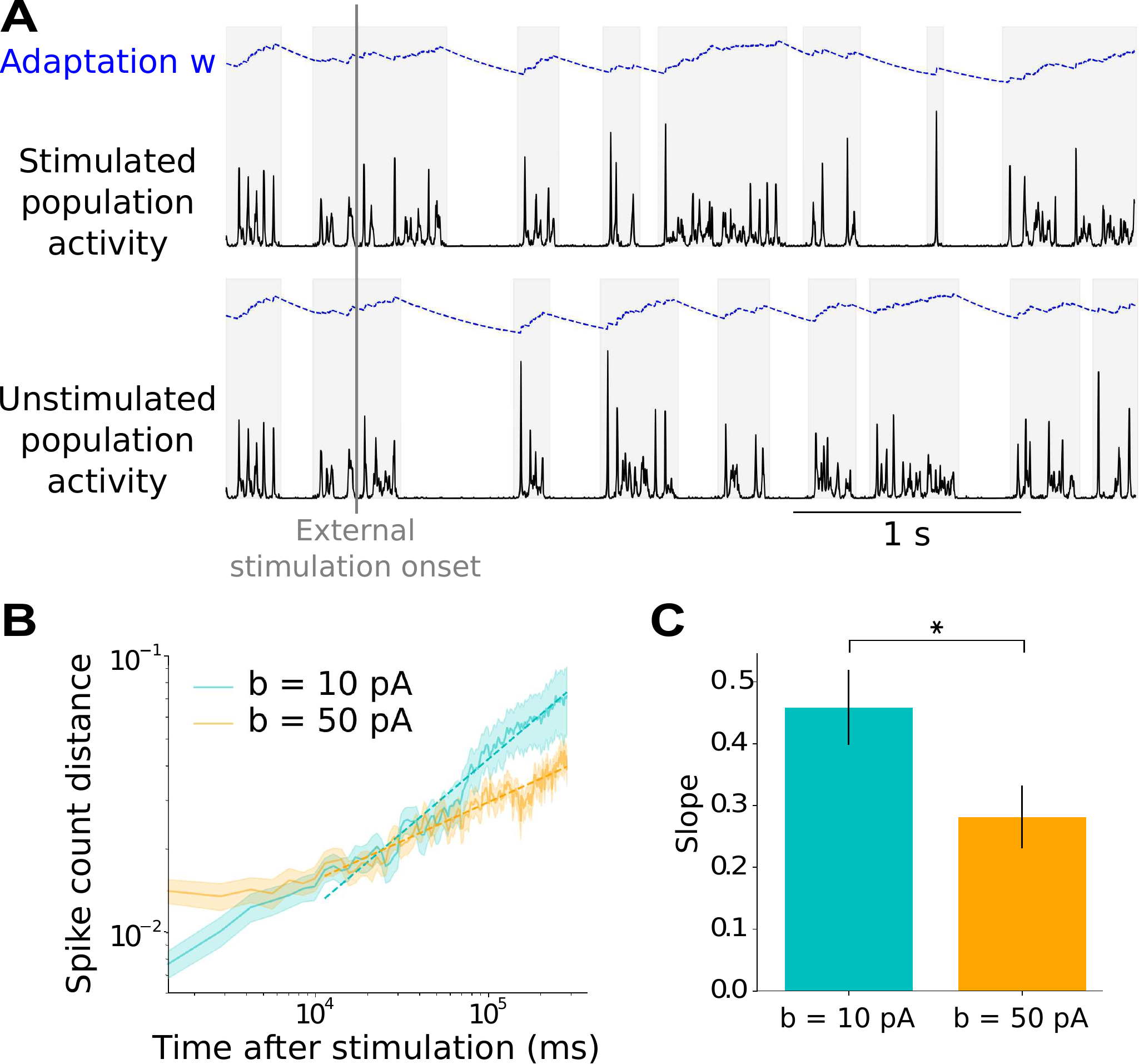}
\caption{
\textbf{Higher sensitivity to perturbations in sleep-like versus anesthesia-like networks.} (A) 
A stimulus is delivered during an UP state to all neurons in the network (upper panel) and the dynamics  is compared to the system without stimulation (lower panel). 
(B) Absolute difference between population spike counts over time in stimulated and non-stimulated networks, normalised by non-stimulated mean spike count, for two values of adaptation strength $b$, averaged over trials (shaded area: standard error of the mean over all trials, see Methods). 
(C) Linear regression slope for all trials, for the two different values of adaptation strength (error bars: standard error of the mean over all trials). The slope is significantly larger (independent Student's T-test, p = 0.03) for lower adaptation, denoting less stable dynamics and increased capacity for sustained stimulus encoding in the sleep-like case.}
\label{fig4_sensitivity}
\end{figure}

To investigate functional consequences of this difference in neuromodulation between sleep and anesthesia, we perform simulations to evaluate to what extent a stimulus affects collective dynamics. As shown in Fig. \ref{fig4_sensitivity}, the stimulus is simulated by delivering a spike train with Poisson statistics to all neurons, and the spike count after the stimulation is compared in the presence and in the absence of the stimulus (see Methods). Immediately after stimulation, the anesthesia-like network is more responsive: as higher adaptation makes the network more silent, more spikes are evoked by the stimulus, relative to spontaneous firing rates. However, in the sleep-like network the difference between stimulated and non-stimulated dynamics diverges significantly faster, which shows lower adaptation makes the dynamics  more sensitive to a stimulus. After tens of seconds, we therefore find that the difference between stimulated and non-stimulated networks is larger for low adaptation, suggesting a more sustained response to the stimulus in the sleep-like case. The findings support that low adaptation strength renders sleep-like networks more sensitive to perturbations, thus allowing enhanced capacity for encoding than in anesthesia-like networks.

An interplay between adaptation strength and noise amplitude can account for global differences in the temporal correlation structure of sleep-like versus anesthesia-like slow waves, but also subtle dynamical differences including greater symmetry of the distribution of correlations in rat (Fig.  ~\ref{fig1_data}F) than human SWS (Fig.  ~\ref{fig1_data}C). Indeed, based on the mechanism here proposed, a possible explanation for this distinction may be different amounts of adaptation and noise in different species and brain regions: stronger adaptation will tend to bias correlations between UP and next DOWN state durations toward positive values, but not DOWN to next UP, resulting in the asymmetry that is observed in the human data. When adaptation strength $b$ is large enough, after long UP states a large amount of adaptation has built up and therefore adaptation $w$ is large. Therefore, it takes time for adaptation to decay, time during which small noise fluctuations do not succeed in destabilising the DOWN state and triggering the next UP state. In other words, for sufficiently strong adaptation compared to noise amplitude levels, long UP states tend to be followed by long DOWN states, as seen in human but not rat sleep in the case of our recordings. 

Notice that adopting the simplest rule for noise amplitude variation, i.e. a simple step wise function, is shown to be enough to account for both types of slow waves. Nevertheless, we verify, by changing the time duration of each step $T$, how the statistics of the process describing noise amplitude dynamics affects more refined observables such as variations in the correlation between state durations with lags (see Fig. ~\ref{fig1_data}C). We find that for sufficiently small values (around 25ms) of $T$, negative correlations decay but remain significant after long lags at low adaptation, reminiscent of observations during sleep (Fig. ~\ref{fig1_data}C,F), while positive correlations are non-significant for non-consecutive states, similar to recordings under anesthesia (Fig. ~\ref{fig1_data}I). While this approach yields satisfactory results a more elaborated study would be necessary in the future to understand the impact of realistic noise processes for the structure of correlations across lags in the two slow waves regimes.

The findings indicate that higher gain for inhibitory than excitatory neurons is a necessary mechanism to model sleep-like slow waves. Transitions from sleep-like to anesthesia-like dynamics can be accounted for by increasing adaptation strength which filters out fluctuations and perturbations. Therefore, the model predicts that anesthesia slow waves are less sensitive to fluctuations and therefore more stationary than sleep slow waves, and that enhancing choninergic neuromodulation to decrease adaptation can lead to a change in slow wave type.

\subsection*{Enhanced stationarity in anesthesia compared to sleep slow waves consistent with model prediction}
A crucial ingredient of our model is the variability $\Delta$ in noise amplitude, especially in the sleep-like regime. Indeed, for lower adaptation strength, such as in SWS, noise fluctuations may play a larger role in shaping UP and DOWN state dynamics. Accordingly, we expect to observe a higher variability in UP and DOWN state durations (as a direct outcome of noise variability, see Fig. \ref{fig1_data}) during sleep with respect to anesthesia. Comparing the mean empirical values of UP and DOWN state durations over relatively long time windows of 100 slow-wave cycles (i.e. $10^2$ s), we observe a higher variability, of order 200\%, in sleep as compared to under anesthesia, as predicted by the model (see Methods).
To further characterize the time scale of noise fluctuations, the time window size was varied, and the correlations across windows were studied. 
By collecting UP and DOWN durations in each window during sleep, we observe that, just as in our model, UP and DOWN state durations belonging to different windows have different correlation values. For short time windows (up to the order of 50 cycles), the Pearson coefficient is positive in the majority of windows, but becomes negative when computed over longer windows (see Supplementary Material). This suggests that fluctuations take place at a time scale $T$ that can be at the fastest of the order of $10$ seconds. Still, the time scale $T$ is relatively slow as compared to that of UP and DOWN dynamics ($\sim 1$ s). This confirms the previous assumption that the time scale $T$ of fluctuations is longer than the UP and DOWN cycle duration (of the order of 1 second).
 Conversely, $T$ is much shorter than the duration of all our recordings (12 minutes to 3 hours) for either sleep or anesthesia, such that the absence of a negative correlation during anesthesia cannot be explained by too short recordings (unless $T$ in anesthesia is not the same as in sleep, but much longer than the duration of the recordings studied here).

Additionally, background fluctuations at time scale $T$ are present during sleep but absent during anesthesia, consistent with the apparent long memory of the UP and DOWN state duration correlation in SWS. Indeed, one may consider a period of time $T$ over which background noise may be approximated as constant. During low noise amplitude periods, UP states are long and DOWN states short; during high noise amplitude periods, UP states are short and DOWN states long. Independent of the lag, UP state durations undere negatively correlated to DOWN state durations proximal in time, provided that they occur within the same period of duration $T$. It is verified that the order of magnitude of $T$ matches that of the time scale of the memory in sleep (order of 10 UP and DOWN cycles, Fig. \ref{fig1_data}C, F, I).

\subsection*{Increasing cholinergic modulation \textit{in vitro} causes transition from anesthesia- to sleep-like dynamics}

Another prediction of our model is the ability of adaptation strength to modulate correlations between UP and DOWN state durations. Spike-frequency adaptation simulates an effective action of activity-dependent potassium conductances, physiologically affected by neuromodulators \cite{mccormick1989convergence}. It has been observed that neuromodulation is depressed during anesthesia \cite{jones2003arousal}, and thus the strength of adaptation should be increased \cite{mccormick1989convergence}. This is consistent with our model prediction, where a transition to anesthesia (higher adaptation) yields a positive correlation between UP and DOWN states.

\begin{figure}[tbhp]
\centering
\includegraphics[width=.85\linewidth]{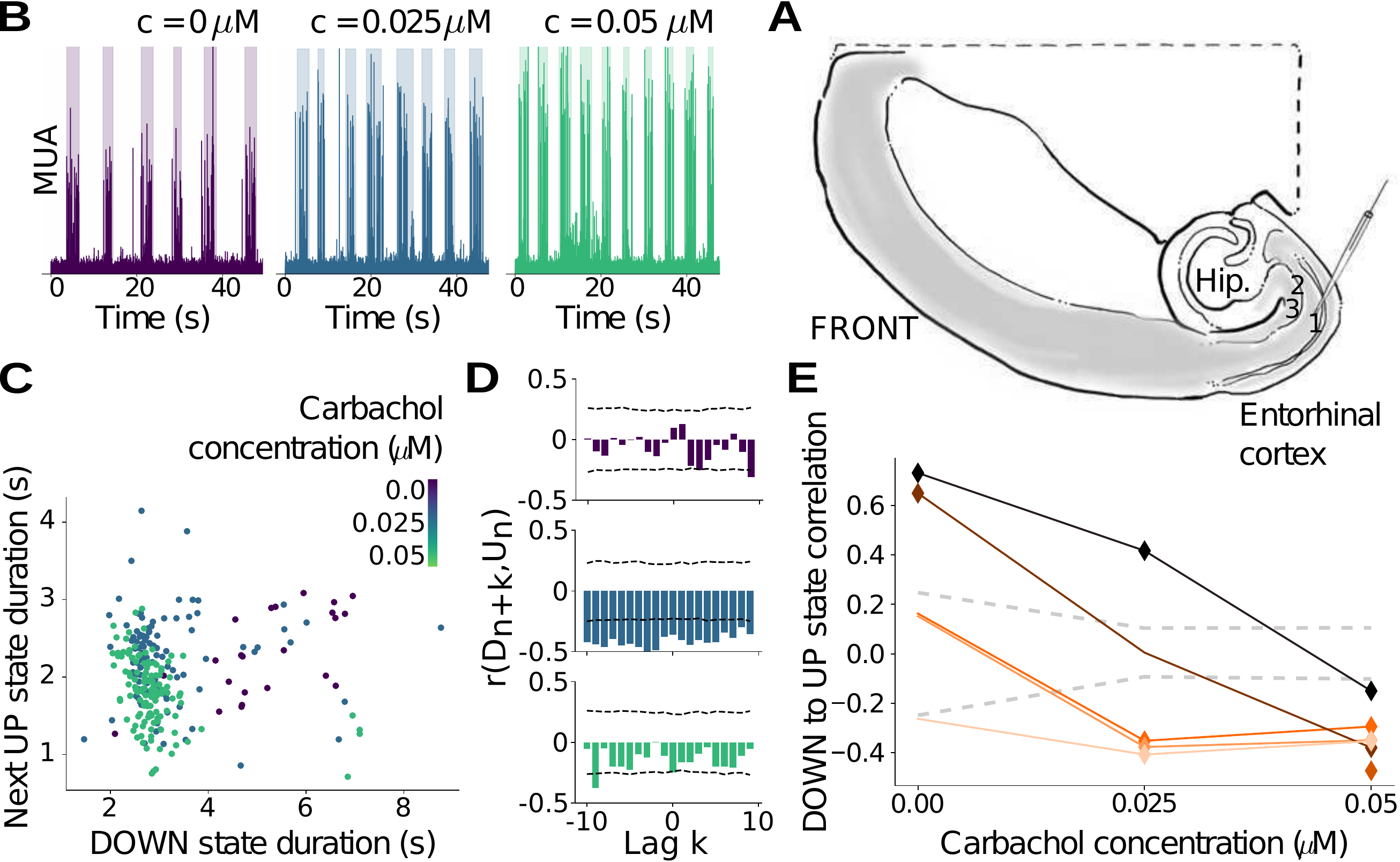}
\caption{\textbf{Reducing adaptation by addition of carbachol in mouse slice preparations produces a transition from anesthesia-like to sleep-like dynamics} (A) Illustration of a horizontal mouse brain slice. Extracellular recording pipettes for Multi-Unit Activity (MUA) were placed in layer 2/3 of the entorhinal cortex. (B) MUA throughout time, recorded for different carbachol concentrations c for an example slice (shaded: UP state detection). UP state frequency is observed to increase with carbachol concentration. (C) For the same example slice, UP state against previous DOWN state duration for different carbachol concentrations, showing a positive correlation (r = 0.64, p $<$ 0.05) in the control condition ($c = 0 \mu M$), non-correlated for an intermediary concentration ($c = 0.025 \mu M$, r = 0.00), and a negative correlation ($c = 0.05 \mu M$, r = -0.37, p $<$ 0.05) when the highest carbachol concentration is added. (D) Bar plot of Pearson correlation $r(D_{n+ k}, U_n)$ as a function of lag $k$ for different carbachol concentrations in an example slice. Two standard deviations of the Pearson correlations when shuffling state durations (dashed lines) provide an interval of confidence outside of which empirical correlations may be considered non-trivial (see Methods). (E) For all recorded slices, correlation between UP state and previous DOWN state duration as a function of carbachol concentration. Consistent with model predictions on the effect of adaptation strength, all slices exhibit a positive or non-significant, `anesthesia-like' correlation in the control condition ($c = 0 \mu M$) and a negative, `sleep-like' correlation for the highest carbachol concentration ($c = 0.05 \mu M$), when adaptation is blocked (markers: significant correlations, colors: different slices, dashed lines: shuffles, see Methods).}
\label{fig5_slice}
\end{figure}


We therefore conducted an experiment to directly validate our prediction.
To this purpose we performed extracellular recordings of neural activity in acute slices of entorhinal cortex  from wild type juvenile mice. A great advantage of \textit{in vitro} preparations is the possibility to pharmacologically modulate adaptation \cite{mccormick1989convergence} by dissolving neuromodulators in the artificial cerebrospinal fluid (ACSF) in which the slice is recorded. To cause an effective decrease of adaptation strength,  carbachol, an agonist of both nicotinic and muscarinic acetylcholine (ACh) receptors is used \cite{mccormick1989convergence}. By increasing concentrations of carbachol (0, 0.025, 0.05 $\mu$M), we observe lengthening UP states and shortening DOWN states, Fig. \ref{fig5_slice}, in accordance with a lower amount of spike-frequency adaptation, as in the model. Importantly, we report a clear tendency to more negatively correlated, sleep-like UP and DOWN state durations at higher carbachol concentrations. 

Additionally, in the presence of intermediate concentrations of carbachol (c = 0.025 $\mu$M), negative correlations are found to tend to persist at long time scales, across multiple slow-wave cycles, consistent with sleep-like dynamics (Fig. ~\ref{fig5_slice}D). However the negative correlation vanishes upon further addition of carbachol (c = 0.05 $\mu$M). A possible explanation is that carbachol does not only reduce adaptation, but also increases neural excitability, which may result in a shorter time scale of noise fluctuations, as more neurons participate, and more frequently, in generating noise fluctuations that trigger or maintain UP states. In turn, faster noise fluctuations lower the time scale at which negative correlations between state durations remain.

In summary, the \textit{in vitro} manipulations using cholinergic agonist carbachol support the model prediction by which decrease of adaptation strength permits a transition from anesthesia-like to sleep-like slow waves.

\section*{Discussion}

In the present paper, we have shown two types of slow waves by analyzing multi-electrode extracellular recordings in human and animal cerebral cortex. This conclusion was reached based on the temporal patterns of UP and DOWN states, an alternation of which causes slow waves to emerge. A first type of slow waves shows positively correlated consecutive UP and DOWN state durations and has been found in previous work \cite{jercog2017up}. Here we report a second type of slow waves displaying negatively correlated UP and DOWN states that remain correlated across longer distances in time. Further, we report that the first type of slow waves is consistently observed in anesthetized states, while the second type is ubiquitous during natural slow-wave sleep. Using computational models, we show that longer UP states correspond to conditions with diminished adaptation, higher noise amplitude, and higher gain on inhibitory than excitatory cells. In contrast, longer DOWN states can be obtained through increasing adaptation or decreasing noise amplitude. 

In summary, our simulations and experiments demonstrate a cholinergic switch between the two types of slow waves distinguished for the first time, to our knowledge, here. We further show that the two types of slow waves differ in their sensitivity to external inputs, demonstrating that perturbations to networks in anaesthetized type-1 networks are less effective than perturbations in slow wave sleep type-2 networks. 

Specifically, a larger effect of noise fluctuations on network dynamics is observed during sleep, yielding a negative, long-memory correlation between UP and DOWN state durations. Conversely, during anesthesia the dynamics is more stable and characterized by short-memory and positive correlation between state durations. 
    Employing a spiking neuron network model revealed that, during anesthesia, fluctuations are filtered out by the modulatory effect of spike-frequency adaptation, and hence produce much smaller effects on collective dynamics. Cholinergic neuromodulators are known to control spike-frequency adaptation, although other neuromodulators also affect adaptation as well as neural conductance and noise sensitivity \cite{McCormick1992}, which are beyond the scope of this paper. 
    The adaptation-gating based mechanism is demonstrated by modulation of spike-frequency adaptation strength in acute cortical slices. Indeed, addition of charbachol, a cholinergic receptor agonist, induces a switch from anesthesia-like to sleep-like slow waves.
    
    While we report dynamical similarities between \textit{in vivo} an \textit{in vitro}, one should remember that mechanisms underlying time scales at play in collective activity may vastly differ. This may explain subtle differences including the time scale at which correlations persist between state durations \textit{in vivo} (Fig. ~\ref{fig1_data}) and in slices (Fig. ~\ref{fig5_slice}D). Indeed, noise sources differ \textit{in vivo} and \textit{in vitro}, with distinct amplitudes and time scales. First, noise amplitude tends to be lower \textit{in vitro}, and therefore likely to fluctuate through time within a smaller range than \textit{in vivo}. Second, due to a smaller number of neurons i.e. noise generators, and reduced (but not non-existent) connectivity between cortical regions and subcortical regions upon slicing, the time scale of noise fluctuations will also differ between \textit{in vivo} and \textit{in vitro}. In particular, recent work \cite{zierenberg2018homeostatic} has investigated in detail how differences in connectivity and noise statistics affected various features of network dynamics \textit{in vivo} and \textit{in vitro}. In addition, certain slow time scales are also at play in the slice that are not present \textit{in vivo}, including cell death which progressively reduces overall excitability throughout the experiment, as well as time taken for carbachol to diffuse after being added by hand and be absorbed by the slice.

    For the sake of simplicity, external noise was chosen to be the only effective source of variability in the model. Nevertheless, this is not the only possible choice: one could also consider variability in other parameters in time, that can be tuned to produce longer UP states and shorter DOWN states, like inhibitory conductance \cite{sanchez2010inhibitory} or even adaptation strength (Fig. 4E). In sum, adaptation filters out the effects of time variation of the system's parameters, in anesthesia but not in sleep. 
    
    Furthermore, we show the implications of a switch in slow wave type for the network's sensitivity to stimuli in time, with sleep-like slow waves associated with greater capacity to respond to perturbations in a time-sustained manner. In other terms, slow wave dynamics are less dependent on previous activity and less able to keep track of past inputs to the system in anesthesia than sleep. This may help explain why key cognitive processes, such as memory consolidation, can take place during sleep \cite{wilson1994reactivation,mehta2007cortico,peyrache2009replay}, while anesthesia causes amnesia and memory impairment \cite{culley2003memory,rudolph2004molecular,timofeev2018sleep}. 
    Our observation of two types of slow-waves suggests that the temporal characteristics of slow-wave dynamics reported here may help explain cognitive differences between anesthetized and sleeping states.  For example
    for memory consolidation to occur, the neocortex should encode information by changing its dynamics upon receiving signals from the senses as well as subcortical regions and the hippocampus. This implies that a non-trivial change in external input should be able to modulate the statistics of cortical activity. We showed (Fig. \ref{fig4_sensitivity}) that in models of anesthesia, unlike in models of sleep, strong adaptation filters out the effects of perturbations on UP and DOWN state dynamics, so that any information encoded in the amplitude of inputs to the neural assembly does not affect the network dynamics, and consequently will not be encoded. 
    
The fact that larger gain of inhibitory cells was necessary to obtain sleep-like dynamics in our model is consistent with experimental findings that inhibitory firing rate rises faster than excitatory firing rate upon injecting input current to cells \cite{nowak2003electrophysiological,contreras2003response}. The higher gain has also been reported to be crucial for reproducing other empirical observations in recent modelling work. In particular, inhibitory cells were endowed with enhanced gain to simulate excitatory and inhbitory firing rates at the onset of UP and DOWN states under anesthesia \cite{jercog2017up}. Additionally, a gain difference between excitatory and inhbitory cells was also needed to model network dynamics and responsiveness to multiple interacting stimuli in the awake state \cite{chemla2019suppressive}.

     Our results may also be considered in the light of intracellular dynamical differences between anesthesia and natural sleep, as investigated with ketamine-xylazine anesthesia in cats \cite{chauvette2011properties} and mice \cite{urbain2019brain}. While the former provides evidence for reduced depolarization, lower slow wave amplitude and longer DOWN states, consistent with stronger adaptation in anesthesia than sleep, the latter reports increased depolarization in anesthesia as compared to sleep, and enhanced synchrony between thalamus and cortical slow waves. By comparing different species, brain regions, and anesthetics, we highlight that despite variations across recordings of the depolarization and subsequent absolute duration of DOWN and UP states, a pattern is robustly followed in terms of correlation between the duration of consecutive states for all experimental conditions studied here. At the opposite end of the spatial scale, it may also be useful to inspect the results in comparison to macroscopic studies having compared activation patterns across brain regions in sleep and anesthesia \cite{brown2010general, AkejuBrown2017,urbain2019brain, chauvette2011properties}. Indeed we provide a possible mechanistic explanation, based on filtering of information transfer between neural assemblies by adaptation, which could lead to differential communication between brain regions during sleep and anesthesia at the macroscopic scale.
    
    In addition to revealing dynamical differences between sleep and anesthesia in spiking activity, we also provide a proof of concept that both slow-wave types can be identified from the LFP. It should be noted, however, that the methods presented here are by no means a definitive approach to detecting UP and DOWN states from LFP.  More work will likely be needed in the future to establish a ground truth to help identify UP and DOWN states from LFP, but this is outside the scope of this paper. Rather, our work supports that the two types of slow waves can be investigated in continuous neural signals such as the LFP, in addition to microscopic (unit) signals. Therefore our approach might prove promising when applied to other continuous and macroscopic brain signals in the future, such as magneto- and electroencephalography.
    
    One should note that stronger fluctuations in slow wave dynamics observed during sleep than anesthesia, at a period around several seconds to tens of seconds (slower than slow-wave cycles), are consistent with previous results in electroencephalography reporting an alternation between distinct oscillatory patterns at a similar time scale of tens of seconds, with autonomic and motor correlates \cite{parrino2012cyclic}. Sources of fluctuations in sleep slow waves within the brain may originate outside neocortex as hippocampal inputs are known to modulate neocortical UP and DOWN state dynamics \cite{battaglia2004hippocampal}, as well as midbrain reticular neurons that discharge with period around 18s \cite{steriade1982firing}, and thalamic spindles with period 5s \cite{nicolas1997four}. 
    
    
    It may also be interesting to consider different types of slow waves in the context of sleep pressure. In particular, at the beginning of a night's sleep, one may hypothesize that higher sleep-pressure could be associated with more anesthesia-like slow waves, with DOWN-to-UP state duration correlations biased toward positive values and only present at short time scales. Indeed, when sleep occurred in ecological conditions, DOWN states were progressively shortened throughout the night as sleep pressure lightened \cite{vyazovskiy2009cortical}, consistent with the hypothesis that cholinergic levels might rise with time spent asleep, leading to a decrease adaptation strength. In the presence of numerous nights of sleep recordings under ecological conditions, investigating sleep pressure in light of our findings may therefore provide insight into the mechanistic underpinnings of sleep pressure and its effects on sensitivity to stimuli and on memory.
    
    Our study also sheds novel light upon current understanding how the dynamics and functional roles of SWS compare to other sleep stages, in particular REM. With respect to subcortical neuromodulation, cholinergic mechanisms are also considered important in REM sleep control \cite{jasper1971acetylcholine, vazquez2001basal}. From a functional point of view, SWS and REM are believed to play different, perhaps complementary roles in memory creation and consolidation processes \cite{vyazovskiy2014nrem}. Specifically, slow-wave dynamics have been proposed in recent work, using anesthesia as a model of SWS, to promote forgetting through homeostatic mechanisms \cite{gonzalez2018activity}, while REM sleep could be responsible for forming new memories \cite{watson2015sleep}. Conversely, we provide evidence suggesting that during sleep-like but not anesthesia-like slow wave dynamics, cholinergic neuromodulation allows cortical networks to encode inputs in a time-sustained fashion, potentially promoting plasticity.
    
    Finally, we emphasize that understanding the role of adaptation in filtering out external variability may shed light on pathological conditions. Indeed, slow oscillations are a hallmark of pathological unconscious states, found in severely brain injured patients and vegetative state patients \cite{sanchez2017shaping}. In such patients, disruption of causality and complexity in response to transcranial magnetic stimulation \cite{rosanova2018sleep} appears consistent with anesthesia-type slow waves, suggesting adaptation-like mechanisms could play a critical role in impairing information representation. Furthermore, deceleration of slow oscillatory patterns during SWS \cite{mander2016sleep,castano2017slow} and loss of memory \cite{prinz1982sleep} are also biomarkers of Alzheimer's disease. The cholinergic system, that modulates spike-frequency adaptation \cite{mccormick1989convergence}, also breaks down in Alzheimer's disease \cite{kihara2004alzheimer}. Our results therefore suggest that the loss of acetylcholine in Alzheimer's disease should increase spike-frequency adaptation, similarly to under anesthesia. It is therefore conceivable that the neocortex of Alzheimer's patients cannot encode fluctuating inputs from the hippocampus during sleep due to anesthesia-like slow-wave dynamics. This mechanism may contribute to explaining why new memories cannot be formed, and to better comprehending how treatments restoring acetylcholine levels alleviate Alzheimer's sleep and memory symptoms \cite{babiloni2013effects}.  Our study directly predicts that the anesthesia-type slow-wave dynamics should be observed in Alzheimer's patients during their natural sleep, and provides an approach to modulating and quantifying the restoration of sleep type slow waves, perhaps a fruitful direction to explore in the design of new therapies.

\makeatletter 
\renewcommand{\thefigure}{S\@arabic\c@figure}
\makeatother

\setcounter{figure}{0} 
\begin{figure}[!htb]
\centering
\includegraphics[width=.55\linewidth]{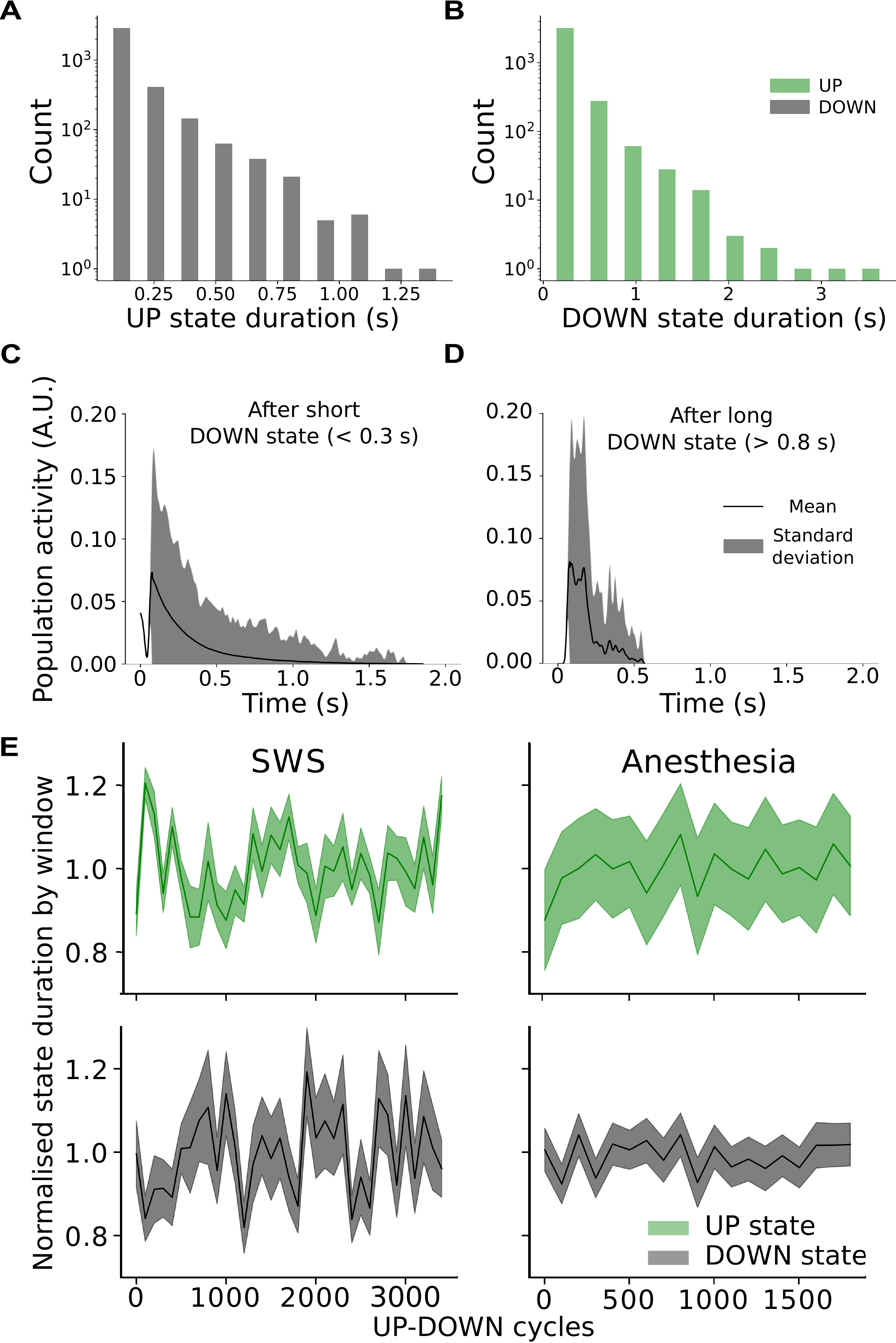}
\caption{\textbf{UP and DOWN state statistics in experimental data.}  (A-D) Slow-wave sleep data from human temporal cortex. (A-B) Both UP and DOWN state durations follow exponential long-tailed, distributions. (C-D) Averaged population firing rate, aligned to UP state onset, after short (C) or long (D) DOWN state durations. UP states following very long DOWN states, $> 0.8 s$, are always short, while UP states following the shortest DOWN states, $< 0.3 s$, can be up to around three times longer.  (E) Mean UP and DOWN state durations in 100 slow-wave-cycle windows in time, normalized by the mean duration over the whole recording, highlight larger fluctuations across windows in sleep (human data, left) than anesthesia (rat data, right). The mean in each window is represented by a full line, while the standard error of the mean is represented by the shaded area.  }
\label{figS1_data}
\end{figure}

\begin{figure}[!htb]
\centering
\includegraphics[width=.95\linewidth]{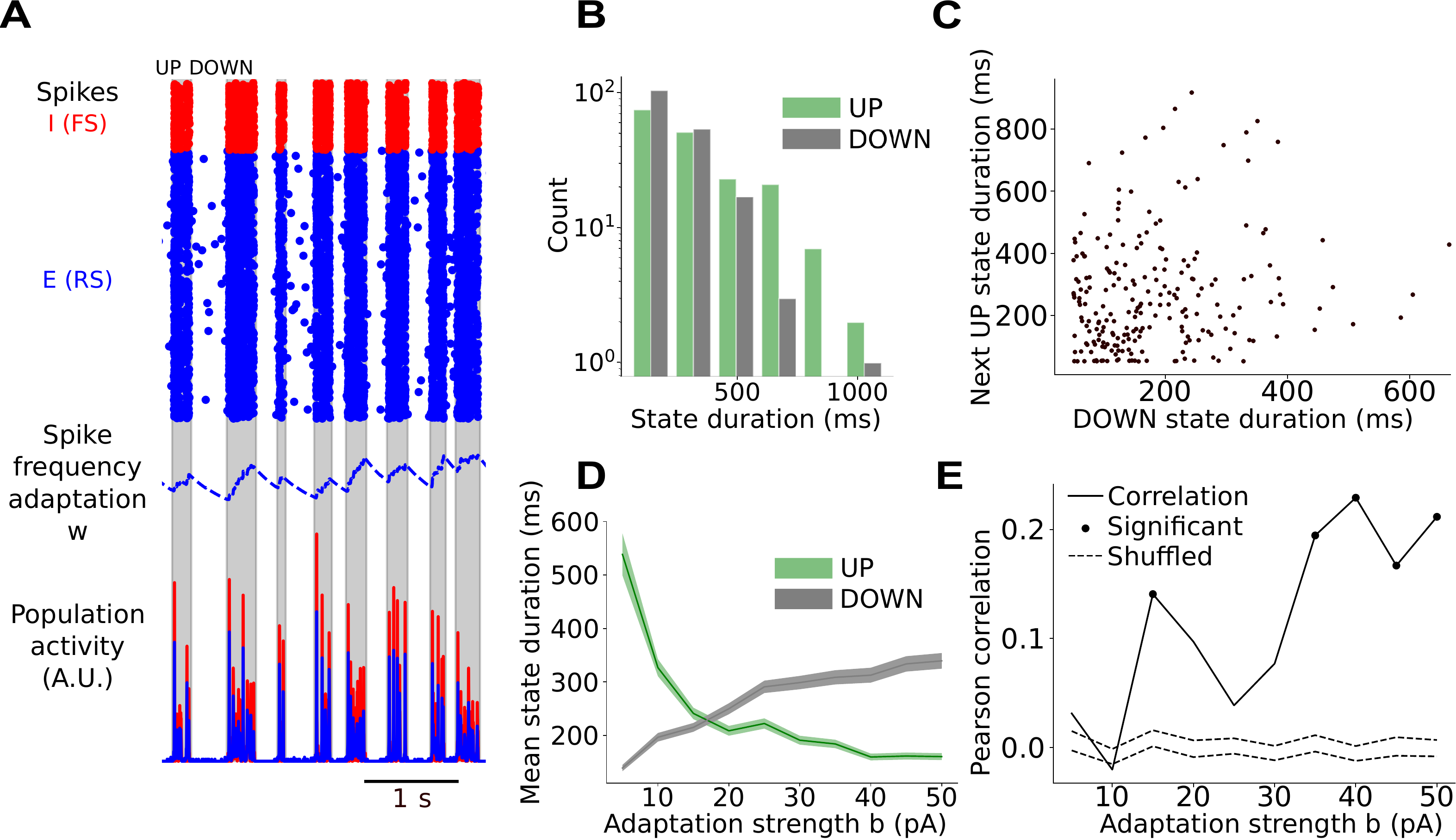}
\caption{\textbf{UP and DOWN states in adaptive exponential integrate-and-fire model.}  (A) Spikes and population activity produced by a spiking model of a RS (blue)-FS (red) neuron network with spike-frequency adaptation on RS cells (blue, dashed line) exhibit UP and DOWN state dynamics (gray: UP state detection). (B) UP and DOWN state durations are exponentially distributed, consistently with empirical data in both sleep and anesthesia. (C) UP state against previous DOWN state durations yield a significant positive correlation (r = 0.16, p $<$ 0.05) in this example simulation.  (D) State durations in different simulations with increasing adaptation strength, showing shortening UP states and lengthening DOWN states (full line: mean, shaded area: standard error in mean). (E) Modeled UP to previous DOWN state duration correlation against adaptation strength reveals systematic positive or non-significant correlations and indicates an increase of the correlation with adaptation level (markers: significant correlations, shaded: interval of confidence obtained by shuffling, see Methods). This is therefore a good model for anesthesia, but not deep sleep.  }
\label{fig1_model}
\end{figure}

\begin{figure}[!htb]
\centering
\includegraphics[width=.95\linewidth]{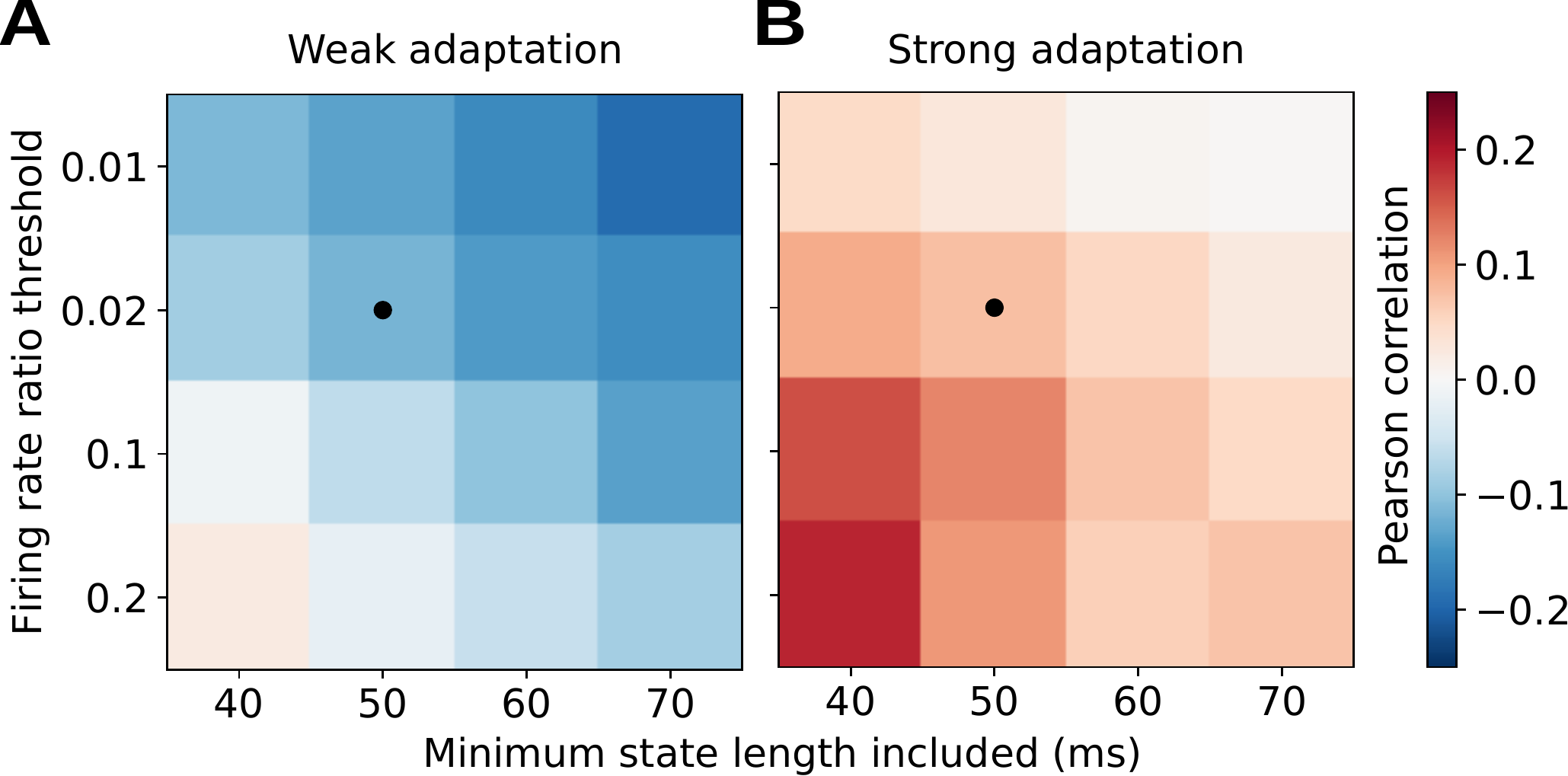}
\caption{\textbf{Pearson Correlation between DOWN and next UP state duration in simulated activity, against different UP and DOWN state detection parameters (minimum state duration and rate threshold for detection, see Methods).}  Shown for (A) low (b = 10 nS) and (B) high (b = 50 nS) adaptation. Black circle: parameters used in the remainder of the paper. Correlations are consistently negative in the low-adaptation, sleep-like system and positive for the high-adaptation, anesthesia-like system, confirming that the observed effect is robust to detection parameters. }
\label{figS3_detection}
\end{figure}

\begin{figure}[!htb]
\centering
\includegraphics[width=.95\linewidth]{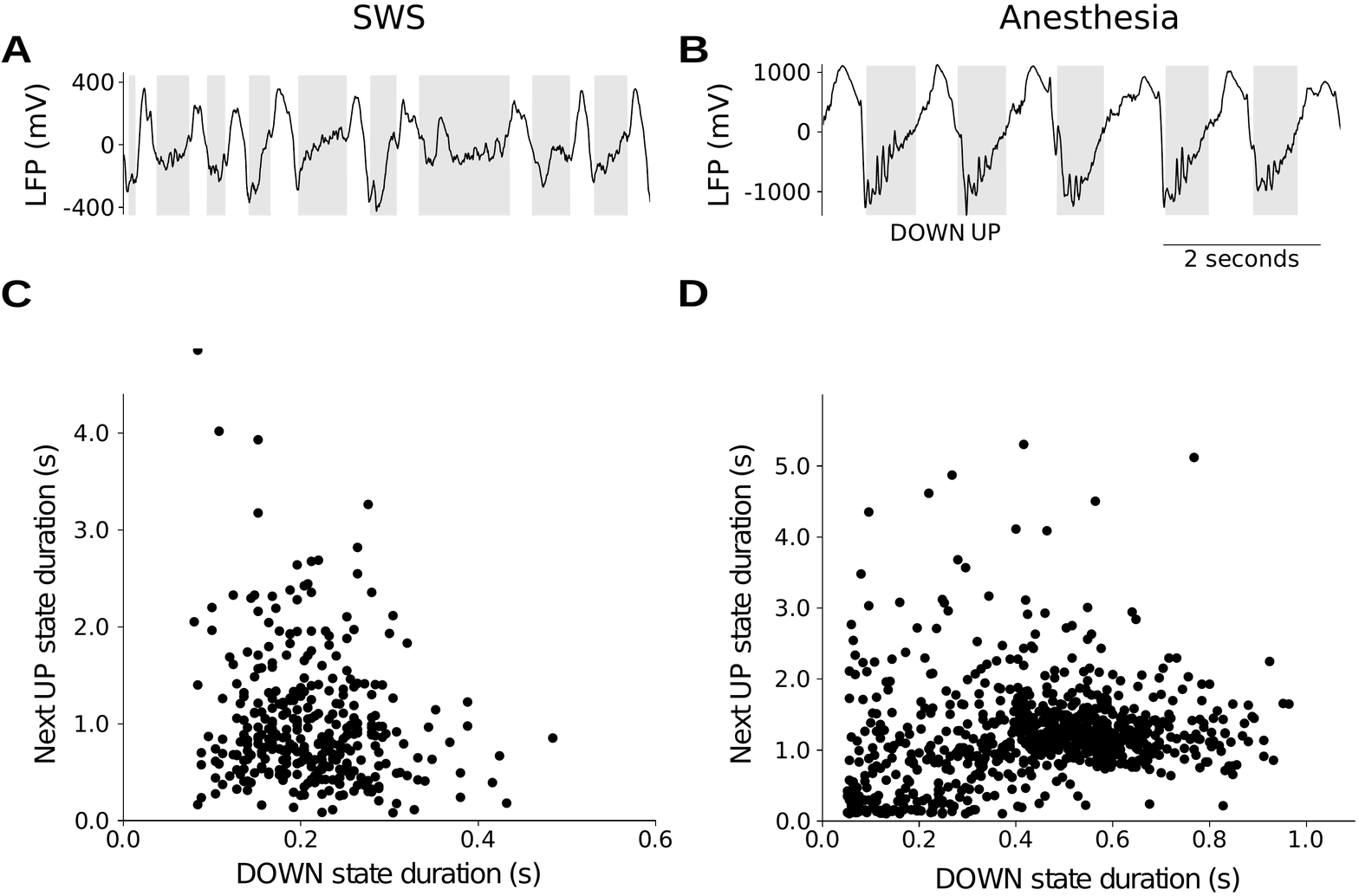}
\caption{\textbf{Slow wave dynamics during sleep and anesthesia in the cat parietal cortex as recorded by LFP.}  LFP signals measured in Brodmann area 5-7 (parietal cortex) during natural sleep (A) and ketamine anesthesia (B) (see methods). Next UP state duration against DOWN state duration in (A) sleep, yielding a negative correlation (r = -0.16, p $<$ 0.01) and (B) anesthesia, yielding a positive correlation (r = 0.16, p $<$ 0.001). The results support that both types of slow waves can be observed within the same species and brain region, and can be studied from LFP.}
\label{figS4_LFPCat}
\end{figure}

\begin{table}[bt]
\caption{\label{tab:example}Model parameters}
\begin{tabular}{|l|l|l|l|}
\hline
{Neuron type} & Parameter name        & Symbol                       & value     \\
\hline
RS \& FS     & Membrane Capacity & $C_m$         & 150pF    \\
RS \& FS     & Leakage Conductance       &$g_L$          & 10nS  \\
RS \& FS     & Excitatory quantal conductance  &$Q_E$            & 1nS    \\
RS \& FS     & Inhibitory quantal conductance & $Q_I$          & 5nS     \\
RS \& FS      & Spiking threshold & $v_{thr}$ &-50mV\\
RS \& FS      & Resting voltage & $v_{rest}$ &-65mV\\
RS \& FS     & Excitatory synapses time decay              & $\tau_E$ &5ms     \\
RS \& FS     & Inhibitory synapses time decay              & $\tau_I$ &5ms     \\
RS \& FS     & Refractory time              & $\tau_r$ &5ms     \\
RS      &Sodium sharpness  & $k_a$                 & 2mV     \\
RS      &Leakage reversal  & $E_L$                 & -60mV     \\
RS     & adaptation current increment   & $b$                   & varies   \\
RS     & adaptation conductance      & $a$                 & 0nS  \\
RS     & adaptation time constant      & $\tau_w$                 & 500ms  \\
FS      &Sodium sharpness  & $k_a$                 & 0.5mV     \\
FS      &Leakage reversal  & $E_L$                 & -65mV     \\
FS     & adaptation current increment   & $b$                   & 0nS   \\
FS     & adaptation conductance      & $a$                 & 0nS  \\
\hline
\end{tabular}
\end{table}

\section*{Acknowledgments}

Research supported by the Centre National de la Recherche Scientifique (CNRS), the Agence Nationale de la Recherche (ANR, PARADOX project) and the European Union (Human Brain
Project H2020-720270 and H2020-785907).
The authors thank D. Contreras, C. Desbois, Z. Giron\'{e}s, M. Mattia, V. Medel, M.V. Sanchez-Vives, and Y. Zerlaut for useful discussion.


\clearpage

\begin{thebibliography}{}

\bibitem[Akeju and Brown, 2017]{AkejuBrown2017}
Akeju O and Brown EN. 2017.
\newblock Neural oscillations demonstrate that general anesthesia and sedative
  states are neurophysiologically distinct from sleep.
\newblock {\em Current opinion in neurobiology}, 44:178--185.

\bibitem[Babiloni et~al., 2013]{babiloni2013effects}
Babiloni C, Del~Percio C, Bordet R, Bourriez JL, Bentivoglio M,
  Payoux P, Derambure P, Dix S, Infarinato F, Lizio R, et~al.  2013.
\newblock Effects of acetylcholinesterase inhibitors and memantine on
  resting-state electroencephalographic rhythms in Alzheimer's disease
  patients.
\newblock {\em Clinical Neurophysiology}, 124(5):837--850.

\bibitem[Battaglia et~al., 2004]{battaglia2004hippocampal}
Battaglia FP, Sutherland GR, and McNaughton BL.  2004.
\newblock Hippocampal sharp wave bursts coincide with neocortical `up-state'
  transitions.
\newblock {\em Learning \& Memory}, 11(6):697--704.

\bibitem[Benchenane et~al., 2010]{benchenane2010coherent}
Benchenane K, Peyrache A, Khamassi M, Tierney PL, Gioanni Y,
  Battaglia FP, and Wiener SI.  2010.
\newblock Coherent theta oscillations and reorganization of spike timing in the
  hippocampal-prefrontal network upon learning.
\newblock {\em Neuron}, 66(6):921--936.

\bibitem[Brette and Gerstner, 2005]{brette2005adaptive}
Brette R and Gerstner W.  2005.
\newblock Adaptive exponential integrate-and-fire model as an effective
  description of neuronal activity.
\newblock {\em Journal of neurophysiology}, 94(5):3637--3642.

\bibitem[Brown et~al., 2010]{brown2010general}
Brown EN, Lydic R, and Schiff ND.  2010.
\newblock General anesthesia, sleep, and coma.
\newblock {\em New England Journal of Medicine}, 363(27):2638--2650.

\bibitem[Bruhn et~al., 2000]{bruhn2000electroencephalogram}
Bruhn J, R{\"o}pcke H, Rehberg B, Bouillon T, and Hoeft A.  2000.
\newblock Electroencephalogram approximate entropy correctly classifies the
  occurrence of burst suppression pattern as increasing anesthetic drug effect.
\newblock {\em Anesthesiology: The Journal of the American Society of
  Anesthesiologists}, 93(4):981--985.

\bibitem[Capone et~al., 2017]{capone2017slow}
Capone C, Rebollo B, Mu{\~n}oz A, Illa X, Del~Giudice P,
  Sanchez-Vives MV, and Mattia M.  2017.
\newblock Slow waves in cortical slices: how spontaneous activity is shaped by
  laminar structure.
\newblock {\em Cerebral Cortex}, pages 1--17.

\bibitem[Castano-Prat et~al., 2017]{castano2017slow}
Castano-Prat P, Perez-Zabalza M, Perez-Mendez L, Escorihuela RM, and
  Sanchez-Vives MV.  2017.
\newblock Slow and fast neocortical oscillations in the senescence-accelerated
  mouse model samp8.
\newblock {\em Frontiers in aging neuroscience}, 9:141.

\bibitem[Chauvette et~al., 2011]{chauvette2011properties}
Chauvette S, Crochet S, Volgushev M, and Timofeev I.  2011.
\newblock Properties of slow oscillation during slow-wave sleep and anesthesia
  in cats.
\newblock {\em Journal of Neuroscience}, 31(42):14998--15008.

\bibitem[Chemla et~al., 2019]{chemla2019suppressive}
Chemla S, Reynaud A, di~Volo M, Zerlaut Y, Perrinet L, Destexhe A,
  and Chavane F.  2019.
\newblock Suppressive traveling waves shape representations of illusory motion
  in primary visual cortex of awake primate.
\newblock {\em Journal of Neuroscience}, 39(22):4282--4298.

\bibitem[Compte et~al., 2003]{compte2003cellular}
Compte A, Sanchez-Vives MV, McCormick DA, and Wang XJ.  2003.
\newblock Cellular and network mechanisms of slow oscillatory activity ($<$ 1 hz)
  and wave propagations in a cortical network model.
\newblock {\em Journal of neurophysiology}, 89(5):2707--2725.

\bibitem[Contreras and Palmer, 2003]{contreras2003response}
Contreras D and Palmer L.  2003.
\newblock Response to contrast of electrophysiologically defined cell classes
  in primary visual cortex.
\newblock {\em Journal of Neuroscience}, 23(17):6936--6945.

\bibitem[Contreras and Steriade, 1995]{contreras1995cellular}
Contreras D and Steriade M. 1995.
\newblock Cellular basis of eeg slow rhythms: a study of dynamic
  corticothalamic relationships.
\newblock {\em Journal of Neuroscience}, 15(1):604--622.

\bibitem[Culley et~al., 2003]{culley2003memory}
Culley DJ, Baxter M, Yukhananov R, and Crosby G.  2003.
\newblock The memory effects of general anesthesia persist for weeks in young
  and aged rats.
\newblock {\em Anesthesia \& Analgesia}, 96(4):1004--1009.

\bibitem[Deco et~al., 2009]{deco2009effective}
Deco G, Mart{\'\i} D, Ledberg A, Reig R, and Sanchez-Vives MV.  2009.
\newblock Effective reduced diffusion-models: a data driven approach to the
  analysis of neuronal dynamics.
\newblock {\em PLoS computational biology}, 5(12):e1000587.

\bibitem[Dehghani et~al., 2012]{dehghani2012avalanche}
Dehghani N, Hatsopoulos NG, Haga ZD, Parker R, Greger B, Halgren
  E, Cash SS, and Destexhe A.  2012.
\newblock Avalanche analysis from multielectrode ensemble recordings in cat,
  monkey, and human cerebral cortex during wakefulness and sleep.
\newblock {\em Frontiers in physiology}, 3:302.

\bibitem[Dehghani et~al., 2016]{dehghani2016dynamic}
Dehghani N, Peyrache A, Telenczuk B, Le~Van~Quyen M, Halgren, E, Cash
  SS, Hatsopoulos NG, and Destexhe A.  2016.
\newblock Dynamic balance of excitation and inhibition in human and monkey
  neocortex.
\newblock {\em Scientific reports}, 6:23176.

\bibitem[Destexhe, 2009]{destexhe2009self}
Destexhe A.  2009.
\newblock Self-sustained asynchronous irregular states and up--down states in
  thalamic, cortical and thalamocortical networks of nonlinear
  integrate-and-fire neurons.
\newblock {\em Journal of computational neuroscience}, 27(3):493.

\bibitem[Gonz{\'a}lez-Rueda et~al., 2018]{gonzalez2018activity}
Gonz{\'a}lez-Rueda A, Pedrosa V, Feord RC, Clopath C, and Paulsen O.
   2018.
\newblock Activity-dependent downscaling of subthreshold synaptic inputs during
  slow-wave-sleep-like activity in vivo.
\newblock {\em Neuron}, 97(6):1244--1252.

\bibitem[Holcman and Tsodyks, 2006]{holcman2006emergence}
Holcman D and Tsodyks M.  2006.
\newblock The emergence of up and down states in cortical networks.
\newblock {\em PLoS computational biology}, 2(3):e23.

\bibitem[Jasper and Tessier, 1971]{jasper1971acetylcholine}
Jasper HH and Tessier J. 1971.
\newblock Acetylcholine liberation from cerebral cortex during paradoxical
  (REM) sleep.
\newblock {\em Science}, 172(3983):601--602.

\bibitem[Jercog et~al., 2017]{jercog2017up}
Jercog D, Roxin A, Bartho P, Luczak A, Compte A, and de~la Rocha J.
   2017.
\newblock UP and DOWN cortical dynamics reflect state transitions in a bistable
  network.
\newblock {\em Elife}, 6:e22425.

\bibitem[Jones, 2003]{jones2003arousal}
Jones BE. 2003.
\newblock Arousal systems.
\newblock {\em Front Biosci}, 8(5):438--51.

\bibitem[Kadir et~al., 2014]{kadir2014high}
Kadir SN, Goodman DF, and Harris KD. 2014.
\newblock High-dimensional cluster analysis with the masked EM algorithm.
\newblock {\em Neural computation}, 26(11):2379--2394.

\bibitem[Kihara and Shimohama, 2004]{kihara2004alzheimer}
Kihara T and Shimohama S. 2004.
\newblock Alzheimer's disease and acetylcholine receptors.
\newblock {\em Acta neurobiologiae experimentalis}, 64(1):99--106.

\bibitem[Kohn and Smith, 2016]{kohn2016utah}
Kohn A and Smith M. 2016.
\newblock Utah array extracellular recordings of spontaneous and visually
  evoked activity from anesthetized macaque primary visual cortex (v1).

\bibitem[Le~Van~Quyen et~al., 2016]{le2016high}
Le~Van~Quyen M, Muller LE, Telenczuk B, Halgren, E, Cash S,
  Hatsopoulos NG, Dehghani N, and Destexhe A. 2016.
\newblock High-frequency oscillations in human and monkey neocortex during the
  wake--sleep cycle.
\newblock {\em Proceedings of the National Academy of Sciences},
  113(33):9363--9368.

\bibitem[Mander et~al., 2016]{mander2016sleep}
Mander BA, Winer JR, Jagust WJ, and Walker MP. 2016.
\newblock Sleep: a novel mechanistic pathway, biomarker, and treatment target
  in the pathology of Alzheimer's disease?
\newblock {\em Trends in neurosciences}, 39(8):552--566.

\bibitem[Mattia et~al., 2016]{mattia2016metastable}
Mattia M, Perez-Zabalza M, Tort-Colet N, and Sanchez-Vives, M. 2016.
\newblock Metastable dynamics underlying the multiscale organization of slow
  oscillations.
\newblock In {\em Society for Neuroscience}, San Diego, California.

\bibitem[Mattia and Sanchez-Vives, 2012]{mattia2012exploring}
Mattia, M. and Sanchez-Vives MV. 2012.
\newblock Exploring the spectrum of dynamical regimes and timescales in
  spontaneous cortical activity.
\newblock {\em Cognitive neurodynamics}, 6(3):239--250.

\bibitem[McCormick, 1992]{McCormick1992}
McCormick DA. 1992.
\newblock Neurotransmitter actions in the thalamus and cerebral cortex and
  their role in neuromodulation of thalamocortical activity.
\newblock {\em Progress in neurobiology}, 39(4):337--388.

\bibitem[McCormick and Williamson, 1989]{mccormick1989convergence}
McCormick DA and Williamson A. 1989.
\newblock Convergence and divergence of neurotransmitter action in human
  cerebral cortex.
\newblock {\em Proceedings of the National Academy of Sciences},
  86(20):8098--8102.

\bibitem[Mehta, 2007]{mehta2007cortico}
Mehta MR. 2007.
\newblock Cortico-hippocampal interaction during UP and DOWN states and memory
  consolidation.
\newblock {\em Nature neuroscience}, 10(1):13.

\bibitem[Nghiem et~al., 2018]{nghiem2018maximum}
Nghiem TA, Telenczuk B, Marre O, Destexhe A, and Ferrari U. 2018.
\newblock Maximum-entropy models reveal the excitatory and inhibitory
  correlation structures in cortical neuronal activity.
\newblock {\em Physical Review E}, 98(1):012402.

\bibitem[Nicolas et~al., 1997]{nicolas1997four}
Nicolas A, Rompr{\'e} S, Dumont M, Laberge L, and Montplaisir J.
  1997.
\newblock Four to five seconds periodicity of sleep spindles in different age
  groups.
\newblock {\em Sleep Res}, 26:31.

\bibitem[Niedermeyer et~al., 1999]{niedermeyer1999burst}
Niedermeyer, E, Sherman DL, Geocadin RJ, Hansen HC, and Hanley
  DF. 1999.
\newblock The burst-suppression electroencephalogram.
\newblock {\em Clinical Electroencephalography}, 30(3):99--105.

\bibitem[Nowak et~al., 2003]{nowak2003electrophysiological}
Nowak LG, Azouz R, Sanchez-Vives MV, Gray CM, and McCormick
  DA. 2003.
\newblock Electrophysiological classes of cat primary visual cortical neurons
  in vivo as revealed by quantitative analyses.
\newblock {\em Journal of neurophysiology}, 89(3):1541--1566.

\bibitem[Parrino et~al., 2012]{parrino2012cyclic}
Parrino L, Ferri R, Bruni O, and Terzano MG. 2012.
\newblock Cyclic alternating pattern (CAP): the marker of sleep instability.
\newblock {\em Sleep medicine reviews}, 16(1):27--45.

\bibitem[Paxinos and Watson, 2004]{PaxinosWatson2004}
Paxinos G and Watson C. 2004.
\newblock {\em The rat brain in stereotaxic coordinates}.
\newblock Elsevier Academic, London.

\bibitem[Peyrache et~al., 2012]{peyrache2012spatiotemporal}
Peyrache A, Dehghani N, Eskandar EN, Madsen JR, Anderson WS,
  Donoghue JA, Hochberg LR, Halgren, E, Cash SS, and Destexhe A.
  2012.
\newblock Spatiotemporal dynamics of neocortical excitation and inhibition
  during human sleep.
\newblock {\em Proceedings of the National Academy of Sciences},
  109(5):1731--1736.

\bibitem[Peyrache et~al., 2009]{peyrache2009replay}
Peyrache A, Khamassi M, Benchenane K, Wiener SI, and Battaglia FP.
  2009.
\newblock Replay of rule-learning related neural patterns in the prefrontal
  cortex during sleep.
\newblock {\em Nature neuroscience}, 12(7):919.

\bibitem[Pospischil et~al., 2008]{pospischil2008minimal}
Pospischil M, Toledo-Rodriguez M, Monier C, Piwkowska Z, Bal T,
  Fr{\'e}gnac Y, Markram H, and Destexhe A. 2008.
\newblock Minimal Hodgkin--Huxley type models for different classes of cortical
  and thalamic neurons.
\newblock {\em Biological cybernetics}, 99(4-5):427--441.

\bibitem[Prinz et~al., 1982]{prinz1982sleep}
Prinz PN, Vitaliano PP, Vitiello MV, Bokan J, Raskind M,
  Peskind E, and Gerber C. 1982.
\newblock Sleep, EEG and mental function changes in senile dementia of the
  alzheimer's type.
\newblock {\em Neurobiology of aging}, 3(4):361--370.

\bibitem[Reig et~al., 2010]{Reig2010}
Reig R, Mattia M, Compte A, Belmonte C, and Sanchez-Vives MV.
  2010.
\newblock Temperature modulation of slow and fast cortical rhythms.
\newblock {\em Journal of neurophysiology}, 103(3):1253--1261.

\bibitem[Renart et~al., 2010]{renart2010asynchronous}
Renart A, De~La~Rocha J, Bartho P, Hollender L, Parga N, Reyes A,
  and Harris KD. 2010.
\newblock The asynchronous state in cortical circuits.
\newblock {\em science}, 327(5965):587--590.

\bibitem[Rosanova et~al., 2018]{rosanova2018sleep}
Rosanova M, Fecchio M, Casarotto S, Sarasso S, Casali A, Pigorini
  A, Comanducci A, Seregni F, Devalle G, Citerio G, et~al. 2018.
\newblock Sleep-like cortical off-periods disrupt causality and complexity in
  the brain of unresponsive wakefulness syndrome patients.
\newblock {\em Nature communications}, 9(1):4427.

\bibitem[Rudolph and Antkowiak, 2004]{rudolph2004molecular}
Rudolph U and Antkowiak B. 2004.
\newblock Molecular and neuronal substrates for general anaesthetics.
\newblock {\em Nature Reviews Neuroscience}, 5(9):709.

\bibitem[Ruiz-Mejias et~al., 2011]{RuizMejias2011}
Ruiz-Mejias M, Ciria-Suarez L, Mattia M, and Sanchez-Vives M. 2011.
\newblock Slow and fast rhythms generated in the cerebral cortex of the
  anesthetized mouse.
\newblock {\em J Neurophysiol}, 106(6):2910--2921.

\bibitem[Sadoc, 2014]{sadocELPHY}
Sadoc G. 2014.
\newblock Elphy software.

\bibitem[Sanchez-Vives et~al., 2017]{sanchez2017shaping}
Sanchez-Vives MV, Massimini M, and Mattia M. 2017.
\newblock Shaping the default activity pattern of the cortical network.
\newblock {\em Neuron}, 94(5):993--1001.

\bibitem[Sanchez-Vives et~al., 2010]{sanchez2010inhibitory}
Sanchez-Vives MV, Mattia M, Compte A, Perez-Zabalza M, Winograd M,
  Descalzo VF, and Reig R. 2010.
\newblock Inhibitory modulation of cortical up states.
\newblock {\em Journal of neurophysiology}, 104(3):1314--1324.

\bibitem[Sanchez-Vives and McCormick, 2000]{sanchez2000cellular}
Sanchez-Vives MV and McCormick DA. 2000.
\newblock Cellular and network mechanisms of rhythmic recurrent activity in
  neocortex.
\newblock {\em Nature neuroscience}, 3(10):1027.

\bibitem[Shoham et~al., 2003]{shoham2003robust}
Shoham S, Fellows MR, and Normann RA. 2003.
\newblock Robust, automatic spike sorting using mixtures of multivariate
  t-distributions.
\newblock {\em Journal of neuroscience methods}, 127(2):111--122.

\bibitem[Smith and Kohn, 2008]{smith2008spatial}
Smith MA and Kohn A. 2008.
\newblock Spatial and temporal scales of neuronal correlation in primary visual
  cortex.
\newblock {\em Journal of Neuroscience}, 28(48):12591--12603.

\bibitem[Steriade et~al., 1993]{steriade1993slow}
Steriade M, Contreras D, Dossi RC, and Nunez A. 1993.
\newblock The slow ($<$ 1 Hz) oscillation in reticular thalamic and
  thalamocortical neurons: scenario of sleep rhythm generation in interacting
  thalamic and neocortical networks.
\newblock {\em Journal of Neuroscience}, 13(8):3284--3299.

\bibitem[Steriade et~al., 1982]{steriade1982firing}
Steriade M, Oakson G, and Ropert N. 1982.
\newblock Firing rates and patterns of midbrain reticular neurons during steady
  and transitional states of the sleep-waking cycle.
\newblock {\em Experimental Brain Research}, 46(1):37--51.

\bibitem[Tahvildari et~al., 2012]{2012-tahvildari/mccormick}
Tahvildari B, W{\"o}lfel M, Duque A, and McCormick DA. 2012.
\newblock Selective functional interactions between excitatory and inhibitory
  cortical neurons and differential contribution to persistent activity of the
  slow oscillation.
\newblock {\em Journal of Neuroscience}, 32(35):12165--12179.

\bibitem[Tavoni et~al., 2017]{tavoni2017functional}
Tavoni G, Ferrari U, Battaglia FP, Cocco S, and Monasson R. 2017.
\newblock Functional coupling networks inferred from prefrontal cortex activity
  show experience-related effective plasticity.
\newblock {\em Network Neuroscience}, 1(3):275--301.

\bibitem[Tele{\'n}czuk et~al., 2017]{telenczuk2017local}
Tele{\'n}czuk B, Dehghani N, Le~Van~Quyen M, Cash SS, Halgren E,
  Hatsopoulos NG, and Destexhe A. 2017.
\newblock Local field potentials primarily reflect inhibitory neuron activity
  in human and monkey cortex.
\newblock {\em Scientific reports}, 7:40211.

\bibitem[Timofeev and Chauvette, 2018]{timofeev2018sleep}
Timofeev I and Chauvette S. 2018.
\newblock Sleep, anesthesia, and plasticity.
\newblock {\em Neuron}, 97(6):1200--1202.

\bibitem[Tort-Colet et~al., 2018]{tort2018bimodality}
Tort-Colet N, Capone C, Mattia M, Sanchez-Vives MV, and Destexhe A. 2018.
\newblock Bimodality of cortical up states and thalamic modulation of up state
  duration: an experimental and computational study.
\newblock In {\em Barcelona Computational, Cognitive and Systems Neuroscience},
  Barcelona, Spain.

\bibitem[Tort-Colet et~al., 2019]{tort2019attractor}
Tort-Colet N, Capone C, Sanchez-Vives MV, and Mattia M. 2019.
\newblock Attractor competition enriches cortical dynamics during awakening
  from anesthesia.
\newblock {\em BioRxiv}, page 517102.

\bibitem[Urbain et~al., 2019]{urbain2019brain}
Urbain N, Fourcaud-Trocm{\'e} N, Laheux S, Salin PA, and Gentet 
  LJ. 2019.
\newblock Brain-state-dependent modulation of neuronal firing and membrane
  potential dynamics in the somatosensory thalamus during natural sleep.
\newblock {\em Cell reports}, 26(6):1443--1457.

\bibitem[Vazquez and Baghdoyan, 2001]{vazquez2001basal}
Vazquez J and Baghdoyan HA. 2001.
\newblock Basal forebrain acetylcholine release during rem sleep is
  significantly greater than during waking.
\newblock {\em American Journal of Physiology-Regulatory, Integrative and
  Comparative Physiology}, 280(2):R598--R601.

\bibitem[Vyazovskiy and Delogu, 2014]{vyazovskiy2014nrem}
Vyazovskiy VV and Delogu A. 2014.
\newblock NREM and REM sleep: complementary roles in recovery after
  wakefulness.
\newblock {\em The Neuroscientist}, 20(3):203--219.

\bibitem[Vyazovskiy et~al., 2009]{vyazovskiy2009cortical}
Vyazovskiy VV, Olcese U, Lazimy YM, Faraguna U, Esser SK,
  Williams JC, Cirelli C, and Tononi G. 2009.
\newblock Cortical firing and sleep homeostasis.
\newblock {\em Neuron}, 63(6):865--878.

\bibitem[Watson and Buzs{\'a}ki, 2015]{watson2015sleep}
Watson BO and Buzs{\'a}ki G. 2015.
\newblock Sleep, memory \& brain rhythms.
\newblock {\em Daedalus}, 144(1):67--82.

\bibitem[Wilson and McNaughton, 1994]{wilson1994reactivation}
Wilson MA and McNaughton BL. 1994.
\newblock Reactivation of hippocampal ensemble memories during sleep.
\newblock {\em Science}, 265(5172):676--679.

\bibitem[Zierenberg et~al., 2018]{zierenberg2018homeostatic}
Zierenberg J, Wilting J, and Priesemann V.  2018.
\newblock Homeostatic plasticity and external input shape neural network
  dynamics.
\newblock {\em Physical Review X}, 8(3):031018.

\end{thebibliography}
\end{document}